\documentclass[fleqn,usenatbib]{mnras}
\usepackage[T1]{fontenc}

\DeclareRobustCommand{\VAN}[3]{#2}
\let\VANthebibliography\thebibliography
\def\thebibliography{\DeclareRobustCommand{\VAN}[3]{##3}\VANthebibliography}

\usepackage{graphicx}	% Including figure files
\usepackage{amsmath}	% Advanced maths commands
\usepackage{amssymb}	% Extra maths symbols
\usepackage{xspace}
\usepackage{braket}
\usepackage{booktabs}
\usepackage{rotating}

\RequirePackage{lineno}
\usepackage[utf8]{inputenc}
\usepackage{ae,aecompl}
\usepackage{comment}
\usepackage{natbib}
\usepackage{float}
\usepackage{subcaption}
\usepackage{rotating,times,pictex,graphicx,latexsym}
\usepackage{color}
\usepackage{threeparttable}
\usepackage{lipsum}% to automatically generate some text
\usepackage{multirow}
\usepackage[all]{hypcap}
\usepackage[title,titletoc]{appendix}
\usepackage[]{hyperref}
\PassOptionsToPackage{pdfpagelabels=false}{hyperref}
\usepackage[T1]{fontenc}
\usepackage{newtxtext,newtxmath}
\newcommand{\Ms}{M\ensuremath{_{\odot}}}

\newcommand{\beq}{\begin{equation}}
\newcommand{\eeq}{\end{equation}}

\newcommand{\mum}{\,\ensuremath{\mu}m\xspace}

\newcommand{\x}{\,\ensuremath{\times}\,}

\newcommand{\lsun}{\mbox{\rm L$_{\odot}$}}

\newcommand{\hii}{\mbox{H~{\sc {ii}}~}}
\newcommand{\hi}{\mbox{H~{\sc i}~}}

\newcommand{\degree}{\mbox{$^{\circ}$}}
\newcommand{\av}{\mbox{$A_\mathrm{V}$~}}

\newcommand{\nht}{\hbox{N$(\mathrm H_2)$}}

\title[Scaling relations at clump scale]{
\center Star formation efficiency and scaling relations in parsec-scale cluster-forming clumps}

\author[Rawat et al.]{
Vineet Rawat,$^{1,2}$\thanks{E-mail: vinitrawat1996@gmail.com}
M. R. Samal,$^{1}$ A. Zavagno,$^{3,4}$ Sami Dib,$^{5}$ Davide Elia,$^{6}$ J. Jose,$^{7}$ D.K. Ojha,$^{8}$\newauthor K. Srivastav$^{1,2}$   
\\
$^{1}$Physical Research Laboratory, Navrangpura, Ahmedabad, Gujarat 380009, India\\
$^{2}$Indian Institute of Technology Gandhinagar Palaj, Gandhinagar 382355, India\\
$^{3}$Aix Marseille Univ, CNRS, CNES, LAM Marseille, France\\
$^{4}$Institut Universitaire de France, 1 rue Descartes, 75005 Paris, France\\
$^{5}$Max-Planck-Institut f\"{u}r Astronomie, K\"{o}nigstuhl 17, 69117, Heidelberg, Germany\\
$^{6}$Istituto di Astrofisica e Planetologia Spaziali, INAF, Via Fosso del Cavaliere 100, I-00133 Roma, Italy\\
$^{7}$Indian Institute of Science Education and Research (IISER) Tirupati, Rami Reddy Nagar, Karakambadi Road, Mangalam (P.O.), Tirupati 517 507, India\\
$^{8}$Department  of  Astronomy  and  Astrophysics,  Tata  Institute  of  Fundamental  Research,  Mumbai  400005, India\\
}
\begin{document}
\label{firstpage}
\pagerange{\pageref{firstpage}--\pageref{lastpage}}
\maketitle
\begin{abstract}
Numerical simulations predict that clumps ($\sim$1 pc) should form stars at high efficiency to produce bound star clusters. 
%To gain a better understanding of the physics responsible for the star and cluster formation, the star formation-gas mass relation needs to be explored at scales of clumps in the Milky Way. 
We conducted a statistical study of 17 nearby cluster-forming clumps to examine the star formation rate and gas mass surface density relations (i.e. $\Sigma_{\rm{SFR}}$ vs. $\Sigma_{\rm{gas}}$) at the clump scale. Using near-infrared point sources and $\it{Herschel}$ dust continuum analysis, we obtained the radius, age, and stellar mass for most clusters in the ranges 0.5$-$1.6 pc, 0.5$-$1.5 Myr, 40$-$500~\Ms, respectively, and also found that they are associated with $\Sigma_{\rm{gas}}$ values ranging from 80$-$600 \Ms~pc$^{-2}$. We obtained the best-fit scaling relations as $\Sigma_{\rm{SFR}}$ $\propto$ $\Sigma_{\rm{gas}}^{1.46}$ and $\Sigma_{\rm{SFR}}$ $\propto$ $(\Sigma_{\rm{gas}}/t_{\rm{ff}})^{0.80}$ for the studied sample of clumps. Comparing our results with existing scaling relations at cloud and extragalactic scales, we found that while the power-law exponent obtained in this work is similar to those found at these scales, the star formation rate surface densities are relatively
higher for similar gas mass surface densities. %We also found less scatteredness and a better correlation in the volumetric star formation scaling relation. 
From this work, we obtained instantaneous median star formation efficiency (SFE) and efficiency per free-fall time ($\epsilon_{\rm{ff}}$) of $\sim$20\% and $\sim$13\%, respectively, for the studied clumps. %The mean free-fall efficiency of clumps found in this work is around 15% and with the slope fixed to unity (ΣSFR ∝ Σgas /��_ff ^1.0 ), it is found to be ∼20%, which is significantly higher than the constant free-fall efficiency reported for nearby molecular clouds. 
 We discuss the cause of the obtained high SFE and  $\epsilon_{\rm{ff}}$ in the studied clumps and also discuss the results in the context of bound cluster formation within molecular clouds. We conclude that
%examine the influence of the local environment on setting the scaling relations of clump. 
our results do not favour a universal scaling law with a constant value of $\epsilon_{\rm{ff}}$ in star-forming systems across different scales.
%Overall, the results of this work suggest that star formation is mostly affected and regulated by the local environment and properties of the gas in the localized regions rather than some global galactic scale process.

\end{abstract}


%%%%%%%%%%%%%%%%%%%%%%%%%%%%%%%%%%%%%%%%%%%%%%%%%%

%%%%%%%%%%%%%%%%% BODY OF PAPER %%%%%%%%%%%%%%%%%%

\section{Introduction}
\label{chap_scaling_laws}

%Although most stars form in clustered environments, however only a few of them remain bound after millions of years of evolution, i.e. at least up to 10$^8$ years \citep{lada_lada_2003}. It suggests that most of the clusters (low-mass) are short-lived.
The processes which regulate the conversion of gas into stars are still least understood in molecular
clouds, and so is the cluster formation process. To understand the formation of stellar clusters, it
is required to follow the sequence of interstellar processes from molecular clouds (MCs) to clumps ($\sim$1 pc) and cores ($\lesssim$ 0.1 pc).
However, in reality, this sequential process of star formation is much more complex, involving
filamentary structures of MCs and the relative role of gravity, turbulence, magnetic
field, and stellar feedback that also varies with the scale size, i.e. from clouds to cores \citep{Krum_2014, kra20, pine22, hacar23, Rawat_2024b}. Along
with the aforementioned factors, the radiation pressure from newly formed stars (acting mostly
on dust) or enhanced thermal pressure from photoionized regions can halt the mass accretion to
the clump or core, thereby affecting their outcome \citep{Krumholz_2019}.
%and may eventually 
%unbound the star cluster that form in a clump by violent gas expulsions \citep{Krumholz_2019}. 
So, inquiring about
the star formation rate (SFR) and how it changes at different stages of cloud evolution and spatial scale is a topic of
interest, as it is essential to develop a complete and universal description of star formation in the
Galaxy. In short, the quest is to understand whether star formation is regulated by some large
galactic scale process or dominated by the local conditions of the star-forming gas material in the
region \citep{Dib_2012, Eden_2015, Zhou_2025}.

Stars form in groups, deeply embedded within dense clumps of molecular gas. The fate of a nascent stellar system, whether to become an expanding association or to remain a bound cluster, is determined by how rapidly and effectively it disperses the gas material of the clump/cloud. This process is controlled primarily by two parameters: the efficiency of star formation and the timescale on which the remaining gas is disrupted. It is suggested that, if the gas removal is rapid relative to the free-fall time, then more than half the mass must be in stars for the cluster to remain bound \citep{Hills_1980}. On the other hand, a sufficiently slow mass loss allows a virialized stellar system to expand adiabatically and remain bound. %The stellar feedback, rapid gas expulsions, violent dynamical interactions, and tidal disruptions are the possible reasons for a cluster to become unbound over time and dissolve \citep{Krumholz_2019}. 
%The massive stellar clusters in the universe or in our own Milky Way, which are still bound by gravity even after multiple free-fall times, must be unique in terms of their star formation histories and initial conditions. By star formation history, we mean the star formation efficiency and the timescale in which the gas is converted into stars, which is still not well understood. Some theories suggest that the disruption effect of stellar feedback becomes less significant in high surface density clouds or in massive clusters \citep[see the review article by][]{Krum_2019}. 
Simulations suggest that the star formation efficiency (SFE) within the cluster-forming region impacts the emergence of a rich and bound cluster \citep{Goodwin_2006, Dib_2011, Dib_2013, gee17, shu17, gud18, Krumholz_2019,li19}.
%preventing its emergence as a bound cluster.
%Observations and simulations suggest that the embedded phase of star clusters lies somewhat between 1 to 3 Myr depending upon their mass, and 
%As discussed in chapter \ref{chap_intro}, bound clusters arise only from regions where star-formation efficiency (SFE) is higher than 30\% \citep{lada03,kru19}. 
In addition, it is also suggested that the primordial gas mass, structure, and density profile %the initial spatial and temporal distribution of stars within cluster-forming regions, including the \textcolor{red}{degree of substructure, central concentration,} and degree of mass segregation, 
also plays a decisive role in massive stars and associated cluster formation \cite[e.g.][]{Bonnell_2006, Parker_2014, Chen_2021}. To investigate how the amount of total gas scales with the total number of stars formed, it is important to quantify the SFR and SFE of a star-forming region and understand their correlations with time scales as well as spatial scales.  %Thus, the prerequisite condition to improve our understanding of the formation of intermediate to massive star clusters is to investigate a sample of young clusters of different ages and masses that have recently formed in massive clouds. 
 %Moreover, as discussed in Chapter \ref{chap_intro}, a high-mass gas assembly with a high SFE is also a possible way of forming an intermediate-to-massive stellar cluster. The SFE generally found in molecular clouds is very low \citep[2$-$6\%;][]{Evans_2009, Lada_2010, Heiderman_2010, Evans_2014}, but it can be high in clumps due to their high density.  

%Star formation is important for the evolution of the galaxy; therefore, it is crucial to understand what governs and regulates the star formation process. 
The connection between the SFR and the gas mass is known as the star formation scaling laws. \cite{Schmidt_1959} established, for the first time, that the SFR density is proportional to the square of the density of the gas. %Later on, the molecular hydrogen gas is also included and further studies were expanded to extragalactic scale (Kennicutt, 1989).
\cite{Kennicutt_1998b} investigated a global Schmidt law by measuring the projected star formation rate surface densities ($\Sigma_{\rm{SFR}}$) and gas mass surface
densities ($\Sigma_{\rm{gas}}$) of 61 normal galaxies using H$\alpha$, \hi, and CO observations. This relation at the extragalactic scale is popularly known as the Kennicutt-Schmidt (KS) relation:
%The relation between star formation rate surface density  and gas mass surface density , i.e. the ``scaling law," is well defined at the extragalactic scale by the Kennicutt-Schmidt (KS) relation \citep[][]{Kennicutt_1998b}:
\begin{equation}
\label{eq:KS}
    \Sigma_{\rm{SFR}} = (2.5 \pm 0.7) \times 10^{-4} \left(\frac{\Sigma_{\rm{gas}}}{1 \rm{\Ms\, pc}^{-2}}\right)^{1.4 \pm 0.15} \times  (\rm{\Ms\, yr^{-1} kpc^{-2}}).
\end{equation}
Although large-scale studies offer crucial insights into the correlation between the overall properties of galaxies and the formation of stars, the gas conversion into stars occurs however at a more localized level, i.e. in MCs, down to much smaller scales (clumps and cores).
%The molecular clouds are the sites where stars form, so it is important to constrain the scaling laws at the cloud scale and within single clouds, which is essential for a deeper understanding of the key processes that govern and regulate the formation of stars. 
Studying the scaling relations in the MCs of our own Milky Way Galaxy offers the advantage of higher resolution than any other extragalactic clouds. Therefore, the clouds and the sub-structures within them, along with the stars, can be better resolved and studied at various scales. The earlier studies on scaling relations for MCs in the Milky Way show higher slopes and $\Sigma_{\rm{SFR}}$ values than the KS relation, as well as broad distribution in the $\Sigma_{\rm{SFR}}$ and $\Sigma_{\rm{gas}}$ plot \citep{Evans_2009, Lada_2010, Heiderman_2010, Kennicutt_2012, Lada_2013, Evans_2014}. For example, \cite{Evans_2009} and \cite{Heiderman_2010} studied the SFR$-$gas mass relation for a sample of nearby low-mass  Galactic molecular clouds using the Spitzer Cores to Disks (c2d) and Gould Belt (GB) survey data, respectively, and found that the SFRs in these Galactic clouds lie almost 20$-$30 times above the KS relation. 
%\textbf{However, it is to be noted that the GB survey mainly samples the low-mass star-forming clouds that are located nearby.} %Heiderman et al. (2010) studied nearby molecular clouds from the c2d survey (Evans et al., 2009) and Gould Belt (GB) survey (Dunham et al., 2013) and found 

%Krumholz et al. (2012)
\cite{Krumholz_2012}
argued that $\Sigma_{\rm{SFR}}$ is better correlated with $\frac{\Sigma_{\rm{gas}}}{t_{\rm{ff}}}$, than $\Sigma_{\rm{gas}}$ itself, where $t_{\rm{ff}}$ is the free-fall time-scale of the considered system. %According to this relation, a fraction of the molecular gas is converted into stars over a free-fall time period. 
%In particular, they predicted a linear relation between $\Sigma_{\rm{SFR}}$ and $\frac{\Sigma_{\rm{gas}}}{t_{\rm{ff}}}$⁠. 
The relation between $\Sigma_{\rm{SFR}}$ and $\frac{\Sigma_{\rm{gas}}}{t_{\rm{ff}}}$ is known as volumetric relation and is expressed as: 

%\cite{Krumholz_2012} suggested a volumetric star formation scaling relation accounting for the scatter in the SFR$-$gas mass relation by arguing that the scatter could be due to variation in local free-fall timescales ($t_{\rm{ff}}$), which is defined as 
\begin{equation}
   \Sigma_{\rm{SFR}} = \epsilon_{\rm{ff}} \frac{\Sigma_{\rm{gas}}}{t_{\rm{ff}}},
\label{eq:vol_relation}   
\end{equation} 
where $\epsilon_{\rm{ff}}$ is known as the star-formation efficiency per free-fall time and is a dimensionless measure of the SFR. \citep{Krumholz_2012} find the best fitted value for $\epsilon_{\rm{ff}}$
to be around 0.01 for data involving molecular clouds in the solar neighbourhood, Local Group galaxies, unresolved disks, and starburst galaxies in the local and high-redshift universe. %Based on the turbulent regulated star-formation processes \cite{Krum_mckee_2005} and this observational results, 
\cite{Krumholz_2012} advocates for a universal star formation law across all scales, in which molecular clouds as well as galaxies as a whole convert their mass into stars at a rate of $\sim$1\% of the mass per free-fall time.

\cite{Evans_2014} tested the volumetric star formation relation for 29 nearby
MCs compiled from the Spitzer c2d and GB survey data, and argued that involving
free-fall time does not reduce the scatter in $\Sigma_{\rm{SFR}}- \Sigma_{\rm{gas}}$ relation. \cite{Evans_2014} suggest that  $\epsilon_{\rm{ff}}$ = 0.01  does not predict the behavior of SFRs on small scales within MCs.
%The reasons like observational biases and the impact of physical factors for the weaker correlation and large scatter in the scaling laws at the molecular cloud scale are discussed in detail in Section \ref{motiv:scaling_laws} of Chapter \ref{chap_intro}. 
\cite{Riwaj_2020, Riwaj_2021} re-investigated the scaling relations in 12 nearby MCs and within single clouds using the Spitzer Extended Solar Neighborhood Archive (SESNA) catalogue \citep{Gutermuth_2019} and found a good correlation between $\Sigma_{\rm{SFR}}$ and $\Sigma_{\rm{gas}}$. The authors also showed that the correlation becomes even stronger by including the free-fall time scale in the relation. %For $\Sigma_{\rm{SFR}}-\Sigma_{\rm{gas}}$ relation, the authors found a mean power-law index of $\sim$2.00 with standard deviation $\sim$0.27. 
%\cite{Riwaj_2020, Riwaj_2021} reduced the observational biases by considering various uncertainties in parameters like total stellar mass and gas mass, and better sampling the protostars from the Spitzer Extended Solar Neighborhood Archive (SESNA) catalogue \citep{Gutermuth_2019}. 
Overall, the power-law exponent of the $\Sigma_{\rm{SFR}} \propto \Sigma_{\rm{gas}}^N$ relation changes from 1.0$-$1.5 at extragalactic scales \citep{Bigiel, Kennicutt_2012, Reyes_2019} to 1.5$-$2.0 at molecular cloud scales \citep{Gutermuth_2011, Lada_2013, Evans_2014, Das_2021, Riwaj_2021}. 
It is worth noting that these scaling relations are derived using different methodologies and data sets (Table \ref{tab:scaling_summary}) and thus are sensitive to adopted methodologies and data quality. For example, recent high-resolution magnetohydrodynamic numerical simulations by \cite{Suin_2024} show that the sensitivity and resolution of observations play a significant role in setting the scaling relations. Simulations also suggest various star formation proxies can systematically over- and under-estimate  the actual instantaneous $\epsilon_{\rm{ff}}$ at different times \citep{gru22}.

%\textbf{However, a recent simulation by \cite{Suin_2024} shows that the presence of stellar feedback, reduces the total SFR of a region and, thus, affects the scaling relations. 
%Also, the sensitivity and resolution of observations play a significant role in setting the scaling relations \citep{Suin_2024}. }  

It is believed that the majority of stars
in a MC, if not all, form in star clusters \citep{lada_lada_2003}.
In MCs, the clumps are the actual sites where clusters form, and cores are the sites where individual stars form. Therefore, it is crucial to examine the behaviour of scaling relations from cloud-to-clump-to-core scales for a deeper understanding of the key processes that govern and regulate the formation of stars. In addition, understanding the SFR$-$gas mass scaling relations over different spatial scales is important for the evolution of the SFRs and SFEs from MCs to cores via clumps. However, very few studies on the scaling relations at the clump scale have been done so far \citep[e.g.][]{Heyer_2016}.
%, Zhou_2024b}. 
%In this regard, it is important to analyze the stellar properties, i.e. the SFR and SFE, of a sample of cluster-forming clumps. %Young clusters, which are at the early stages of star formation, tell about the local environment of their parental clump. 
%Also, studying a sample of young clusters would help to delineate the relation between stars and the star-forming material at the clump scale in order to better understand the cluster formation process. %We intend to explore whether the star formation$-$gas mass relation obtained by \cite{Riwaj_2020, Riwaj_2021} will also follow at the clump scale or if there is scope for some variations in the relations. 
%So far, mostly different relations have been found for different data sets depending upon the data quality, scale size, sample size, and observational constraints, but the local gas environment is also believed to play a substantial role. 
In this work, we extend the studies on star formation scaling relations to clump scale by measuring the star formation properties of a sample of clumps hosting embedded clusters (see Section \ref{sec_data}) that are visible in near-infrared (NIR). We then 
discuss the obtained results in the context of star and
star cluster formation in molecular clouds.
%We aim to estimate the gas properties of the clumps and star formation properties  of the clusters formed within them, and then explore their correlations. %To do so, we used near-infrared (NIR) data from the UKIRT Infrared Deep Sky Survey \citep[UKIDSS;][]{lawrence07} and followed a similar methodology as discussed in \cite{Rawat_2024c}. %the properties of the FSR 655 cluster were studied in detail, like its mass, age, extinction, field contamination, SFR and SFE. For the statistical work on studying the SFR$-$gas mass relation at the clump scale, a similar methodology is adopted here that has been used to study FSR 655.       
This work is organized as follows: The data sets and sample selection are presented in Section \ref{sec_data}. The results are presented and discussed in Section \ref{results}. In Section \ref{discuss_1}, we discuss
the obtained scaling relations, as well as the SFE and SFR, with respect to the values obtained over various scales, and we also discuss the results in the context of cluster formation. In Section \ref{sec:summary}, we summarize our work with concluding remarks.

\begin{table*}
\caption{Summary of scaling relations.}
\begin{tabular}{|p{3.8cm} p{3cm} p{2.5cm} p{2.5cm} p{1.5cm} |} 
\hline
\hline

Reference & Scale size (pc) & SFR tracer & Gas mass tracer & $\Sigma_{\rm{SFR}} \propto \Sigma_{\rm{gas}}^N$; $N$ \\ 
\hline
\cite{Kennicutt_1998b} & extragalactic & H$\alpha$ & \hi $+$ CO  & 1.40 $\pm$ 0.15 \\
\cite{Bigiel} 
    & extragalactic ($\sim$750) 
    & \multirow{2}{2.5cm}{$\it{Spitzer}$ mid-IR and FUV maps} 
    & \hi $+$ CO & 1.0 $\pm$ 0.2\\ 
    & & & & \\ % Empty row for alignment
\cite{Evans_2009} & 3$-$10 ($\sim$6)$^{\rm{a}}$ & $\it{Spitzer}$ c2d & Extinction maps (c2d and 2MASS; 240\arcsec) & $\quad$----- \\
\cite{Heiderman_2010} & 1.75$-$15 ($\sim$5.4)  & $\it{Spitzer}$ c2d and GB & Extinction maps (c2d and 2MASS; 270\arcsec) & $\quad$----- \\
\cite{Heiderman_2010} (clumps) & 0.26$-$11 ($\sim$2.26)  
    & Total IR luminosity & HCN (J=1$-$0)  & 1.0  \\
\cite{Lada_2010} & 6$-$66 ($\sim$28) & $\it{Spitzer}$ c2d & \multirow{2}{2.5cm}{Extinction maps (2MASS)} & $\quad$----- \\
& & & & \\ % Empty row for alignment
\cite{Evans_2014} & 1$-$15 ($\sim$4.3) & $\it{Spitzer}$ c2d and GB & Extinction maps (c2d and 2MASS; 270\arcsec)  & $\quad$----- \\
\cite{Heyer_2016} & 0.15$-$10.5 ($\sim$1.5)  & \multirow{2}{2.5cm}{$\it{Spitzer}$ MIPSGAL survey 24$\mu$m} & ATLASGAL 870 $\mu$m and NH$_3$ line emission& $\quad$----- \\
& & & & \\ % Empty row for alignment
\cite{Riwaj_2021} & 11.5$\times$12.0$-$142.2$\times$163.7  & SESNA catalog & $\it{Herschel}$ maps & 2.0 \\
\cite{Elia_2025} & 2000 (Milky Way)   & 70 $\mu$m luminosity (Hi-GAL survey) & CO & 1.10 $\pm$ 0.06 \\

\hline
\multicolumn{5}{|p{13.9cm}|}{\footnotesize Note: $\rm{^a}$ The values quoted in the brackets are the mean scale sizes of the studied regions.} \\ 
\hline
\end{tabular}
\label{tab:scaling_summary}
\end{table*}

\section{Data used and sample selection}
\label{sec_data}

%\subsection{Near-infrared data}

To determine the properties of embedded clusters, NIR photometric data ($J$, $H$, and $K$) from the UKIDSS's Galactic Plane Survey \citep[GPS;][]{lucas2008} of DR10 was used in this work. This survey was done using the Wide Field Camera \citep[WFCAM;][]{Casali_2007} mounted on the United Kingdom Infra-Red Telescope (UKIRT) 3.8-m telescope. The WFCAM has 4 Rockwell Hawaii-II 2048 $\times$ 2048 HgCdTe detectors with a pixel scale of $\sim$0.4\arcsec, and have broadband $ZYJHK$ filters and narrowband H2 and Br$\gamma$ filters \citep{Casali_2007}. %The survey employs the WFCAM instrument mounted on the 3.8-meter UKIRT, which has a field of view spanning 0.21 square degrees with a pixel size of 0.4 arcsec \citep{lawrence07}. 
%The survey area of the GPS includes 1868 square degrees of the northern and equatorial Galactic plane at Galactic latitudes $-$5$\degree$ $<$ b $<$ 5$\degree$ and $\sim$200 square degrees area of the Taurus-Auriga-Perseus molecular cloud complex in the $J$, $H$, and $K$ filters \citep{lucas2008}. 
The GPS survey has a resolution of around 1 arcsec and a typical depth of 19.8, 19.0, and 18.1 mags in the $J$, $H$, and $K$ bands, respectively. %The GPS survey has a resolution of around 1 arcsec. %The GCS survey covers an area of 1067 square degrees that includes 10 open clusters and stellar associations \citep{lawrence07}. 
%The depth of the GPS in $J$, $H$, and $K$ bands is 19.8, 19.0, and 18.1 mags, respectively. %while for GCS, the depth is 19.6, 18.8, and 18.2 mags, respectively. 
In this work, only those point sources were used which have an error of less than 0.1 mag in all three bands. For sources brighter than the saturation limits of the UKIDSS GPS data, i.e. $J$ = 13.25, $H$ = 12.75, and $K$ = 12.0 mag \citep{lucas2008}, the photometric values from the Two Micron All-Sky Survey (2MASS) NIR data are used \citep{Cutri_2003}.

%To study the relation between star formation rate and gas mass, 17 compact clusters that are within a distance of 2.5 kpc were selected. 

Since for measuring the SFR and SFE, both stellar mass and gas mass are to be estimated, we
thus selected clusters that are visible in NIR and are also associated with cold dust. In NIR, although 2MASS has provided a point source catalogue over the whole sky, deep NIR GPS survey, covers mostly the Galactic plane. Similarly, sensitive cold dust continuum images are available for the Galactic plane from the Herschel Infrared Galactic
Plane Survey \citep[Hi-GAL;][]{Mol_2010}. Therefore, in order to adopt a homogeneous approach for estimating stellar and dust properties, we restrict our sample to the Galactic plane within $-2\degree < $ b $ <  2\degree$ encompassing these surveys. We also restrict our sample in the longitude range, $60^\circ <$ l $< 240^\circ$, to avoid sources in the Galactic bulge direction due to high field star densities and high interstellar extinction that would affect the determination of the cluster properties. 
We search for the embedded cluster sample by simultaneously looking at NIR and far-IR Galactic plane images using the ALADIN software \citep{aladin_2022}. While searching, we also consider that the clusters should not be part of a highly structured dusty environment for better estimation of their boundaries and, thus, their properties. Therefore, our sample is biased to those clusters that are largely situated in isolated cold clumpy structures, although in some cases, the clusters are found to be connected to extended structures in their outskirts. In addition, we restrict the cluster sample to be within 2.5 kpc, which is primarily based on the mass sensitivity limit of the GPS survey.  We aimed to detect point sources down to at least 0.5 \Ms~at 2.5 kpc with the typical data completeness of the
GPS survey, so that we can estimate the total stellar mass of the cluster by extrapolating to lower masses (i.e. $\sim$0.1 \Ms) using the functional form of the initial mass function (IMF). %Given that the typical median visual extinctions of a parsec or a few parsec-size young clusters (age $\sim$0.5 to 2 Myr) are in the range of 3 to 15 mag \citep[see Fig. 9 of][]{Rawat_2024c}, we estimated that we would be able to detect 0.5 \Ms~sources up to 2.5 kpc with \av=20 mag with the typical data completeness of the GPS survey \citep{lucas2008} at K-band for clusters as young as 0.5 Myr. 
Given that the typical median visual extinctions of young clusters, ranging from a parsec to a few parsecs in size and aged $\sim$0.5 to 2 Myr, fall within 3 to 15 mag \citep[see Fig. 9 of][]{Rawat_2024c}, we estimate that with the typical data completeness of the GPS survey \citep{lucas2008} at $K$-band, we would be able to detect 0.5 \Ms~sources up to 2.5 kpc. This detection limit holds even in regions with \av= 20 mag for clusters as young as 0.5 Myr.
Since the clusters are associated with cold dust in the $\it{Herschel}$ SPIRE bands and are also either barely visible or invisible in optical images such as DSS-2 and Pan-STARRS, most of the clusters in our sample are expected to be young. Thus, our initial estimate of an age of 0.5$-$2 Myr seems to be a reasonable assumption. We discuss more on the age estimation in Section \ref{KLF_age}. With the aforementioned selection criteria, our final sample consists of 17 clusters in the distance range of 1.5 to 2.2 kpc, which are listed in Table \ref{tab:clusters}. %\textcolor{blue}{We wouldn't like to discuss the outer galaxy and metallically in the paper; this will make the paper more complex and will add one more uncertainty to scaling relations. In the literature, metallically is so far not accounted for, so it will also make one-to-one comparison even worse.}

For estimating the gas mass of the clumps (see Section \ref{clump_gas}), we used the $\it{Herschel}$ column density (\nht) maps of the clumps made with the PPMAP technique by \cite{Marsh_2019}. The PPMAP-based column density maps are constructed using the far-infrared images of the Hi-GAL survey and are available\footnote{\href {http://www.astro.cardiff.ac.uk/research/ViaLactea/PPMAP\_Results/}{http://www.astro.cardiff.ac.uk/research/ViaLactea/PPMAP\_Results/}} for the entire Galactic plane within a strip of around 2 degrees in latitude. The Hi-GAL data is taken with the $\it{Herschel}$ PACS and SPIRE instruments at wavelengths 70, 160, 250, 350, and 500 \mum with angular resolutions of 8.5, 13.5, 18.2, 24.9, and 36.3\arcsec, respectively \citep{Mol_2010}. The PPMAP provides the resolution of 12\arcsec~that corresponds to the spatial resolution of $\sim$0.08$-$0.13 pc for the clumps studied in this work.

\begin{table}
\caption{Sample of clusters.}
%\begin{tabular}{|p{0.08cm} p{2.35cm} p{0.85cm} p{0.85cm} p{1.38cm}  p{0.2cm}|} 

\begin{tabular}{|p{0.08cm} p{2.35cm} r r p{1.38cm} p{0.23cm}|} 
\hline
\hline

No & Name & GLON & GLAT & $D$  & Ref.\\ 
   &      & (degree) & (degree) & (kpc)  & \\
\hline
1 & IRAS 06063+2040 & 189.859 & 0.502 & 2.10 $\pm$ 0.42  & 1\\ 
2 & IRAS 06055+2039 & 189.769 & 0.336 & 2.10 $\pm$ 0.42  & 1\\
3 & IRAS 06068+2030 & 190.053 & 0.538 & 2.10 $\pm$ 0.42  & 1\\
4 & IRAS 22134+5834 & 103.875 & 1.856 & 1.50 $\pm$ 0.30$^*$  & 2\\
%5 & AFGL 5142 & 174.196 & -0.076 & 1.6 & 1\\
5 & IRAS 06056+2131 & 189.030 & 0.784 & 1.76 $\pm$ 0.35 & 1\\
6 & Sh2-255 IR & 192.601 & -0.047 & 1.96 $\pm$ 0.12  & 3\\
7 & [IBP2002] CC14 & 173.503 & -0.060 & 1.80 $\pm$ 0.36  & 1\\
8 & IRAS 06058+2138 & 188.949 & 0.888 & 1.76 $\pm$ 0.35  & 1\\
9 & IRAS 06065+2124 & 189.232 & 0.895 & 2.00 $\pm$ 0.40$^*$  & 4\\
10 & NGC 2282 & 211.239 & -0.421 & 1.70 $\pm$ 0.40 & 5\\
11 & IRAS 05490+2658 & 182.416 & 0.247 & 2.20 $\pm$ 0.44  & 1\\
12 & BFS 56 IR & 217.373 & -0.080 & 2.10 $\pm$ 0.42$^*$ & 6\\
%13 & [BDS2003] 89 & 217.634 & -0.177 & 1.4 & 10.11 & 6\\
13 & Sh2-88 & 61.472 & 0.095 & 2.06 $\pm$ 0.08  & 3\\
14 & IRAS 06103+1523 & 194.931 & -1.210 & 2.00 $\pm$ 0.40$^*$  & 2\\
15 & IRAS 06104+1524A & 194.926 & -1.194 & 2.00 $\pm$ 0.40$^*$ & 2\\
16 & IRAS 06117+1901 & 191.916 & 0.822 & 2.20 $\pm$ 0.44 & 1\\
17 & IRAS 05480+2545 & 183.348 & -0.576 & 2.10 $\pm$ 0.42$^*$ & 7\\
\hline
\hline

\end{tabular}
\begin{tablenotes}[para]
\setlength{\parindent}{0cm}
\hangindent=0cm 
References: [1] \cite{meg21}, [2] \cite{Maud_2015}, [3] \cite{Delgado_2022},  [4] \cite{Dutra_2001}, [5] \cite{Dutta_2015}, [6] \cite{elia_2013}, and [7] \cite{Henning_1992}. Note: $^*$ For clusters without reported distance errors in the literature, we have assumed an average uncertainty of around 20\%. 

\end{tablenotes}

\label{tab:clusters}

\end{table}

\section{Analysis and results}
\label{results}
The present-day SFE is defined as the ratio of the total stellar mass ($M_*$) to the total mass of a star-forming
region, i.e. stellar mass plus present-day gas mass ($M_*$ $+$ $M_{\rm{gas}}$). The SFR is calculated as, $M_*/t_{\rm{clust}}$, where $t_{\rm{clust}}$ is the age of the cluster. The identification and counting of young stellar objects (YSOs) and using the information of their average mass and lifetimes is a direct way to quantify the SFR, which is known as the star-count method. In most of the previous studies, the star-count method was mostly employed for nearby clouds ($<$ 1 kpc) \citep[e.g.;][]{Evans_2009, Heiderman_2010, Lada_2010} due to the low sensitivity of the $\it{Spitzer}$ data at larger distances.  It is suggested that employing average mass and time over molecular cloud scale may not be entirely accurate when star formation is not uniform in time \citep{Dib_2025}. For distant star-forming regions, indirect tracers of the SFR are used like H$\alpha$ emission, ultraviolet continuum, infrared luminosities, and radio continuum emission \citep[see the review article by][]{Kennicutt_1998a, Kennicutt_2012}. %In this study, the UKIDSS NIR data is used to reach a fairly low mass limit \textcolor{red}{($\sim$0.2$-$0.3 \Ms)} for each cluster.  
%To determine the total stellar mass and age of the clusters, we have followed a similar method and steps as discussed in \cite{Rawat_2024c}.

\begin{figure*}
    \centering
    \includegraphics[width=14 cm]{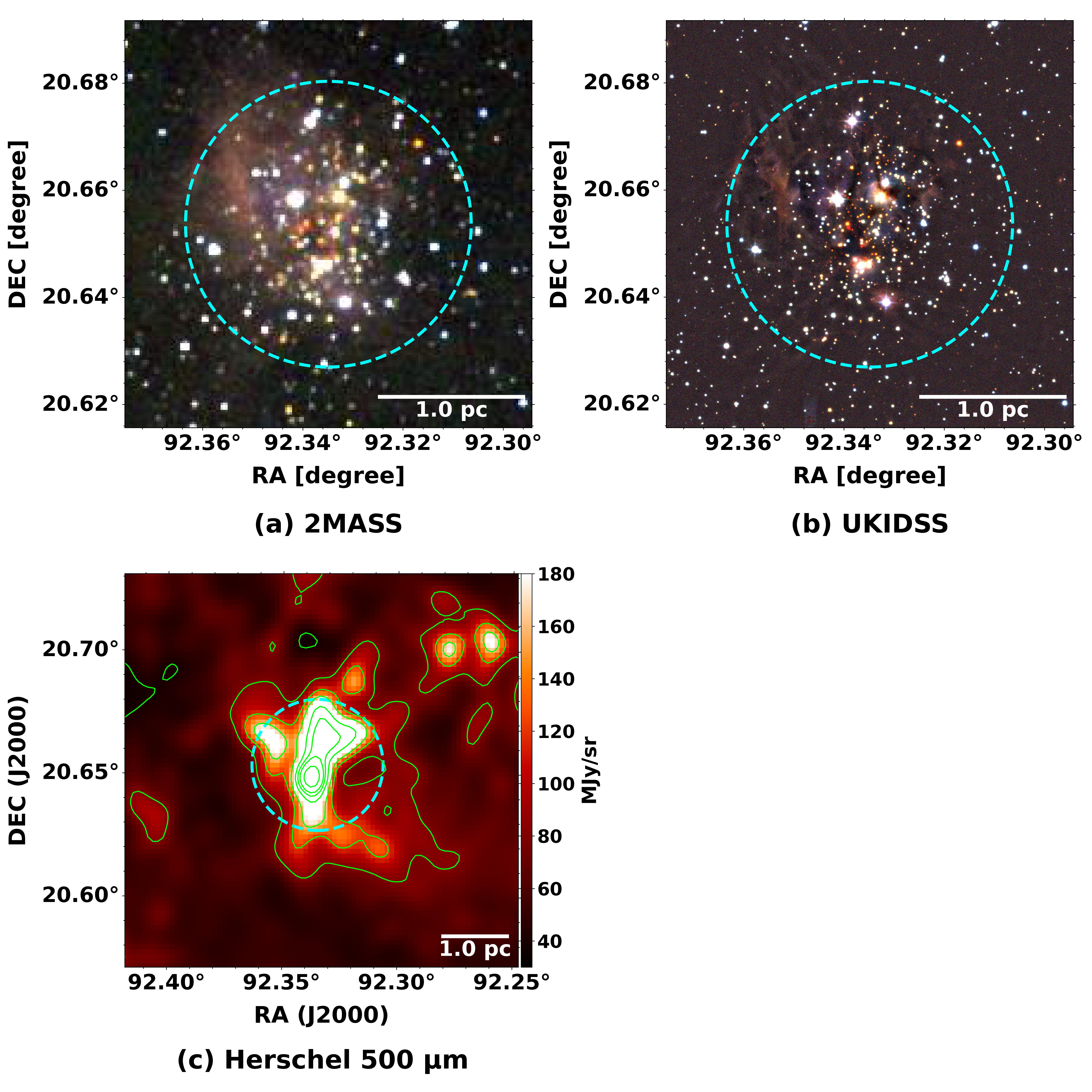}
    \caption{(a) 2MASS and (b) UKIDSS NIR color-composite (Red: $K$ or $K_s$ band for 2MASS; Green: $H$ band, and Blue: $J$ band) image of IRAS 06063$+$2040. The cyan dashed circle shows the extent of the cluster (see Section \ref{cluster_rad}). (c) The $\it{Herschel}$ 500 $\mu$m image of IRAS 06063$+$2040 along with contour levels at 20, 40, 80, 120, 160, 240, 320, 400, and 600 MJy/sr.}
    \label{fig:iras_06063+2040_image}
\end{figure*}

In this work, we explore the star formation properties of a sample of cluster-forming regions of typical size around 1 pc. 
Below, we present the analysis steps that we applied to all the clusters in our sample by showcasing the steps for an exemplary cluster, i.e. IRAS 06063$+$2040.
%and present the results by giving an example of the IRAS 06063$+$2040 cluster and its corresponding plots. 
Fig. \ref{fig:iras_06063+2040_image}a-b shows the 3-color RGB image of IRAS 06063$+$2040 in $JHK$ bands from 2MASS and UKIDSS. Fig. \ref{fig:iras_06063+2040_image}c shows the $\it{Herschel}$ 500~\mum image of IRAS 06063$+$2040, showing the presence of cold dust in the cluster region. As can be seen, point sources in the UKIDSS images are more resolved and richer compared to the 2MASS images, and the distribution of 500 $\mu$m emission around the cluster suggests that the cluster is associated with cold dust.
%, and thus may still be forming stars.

%\subsection{Completeness of the NIR photometric data}
 
%\begin{figure}[h!]
    %\centering
    %\includegraphics[width=14cm]{Chap7_images/IRAS_06063+2040_completeness.jpeg}
    %\caption{Density plots of photometric data of IRAS 06063$+$2040 in $J$, $H$, and $K$ UKIDSS bands. The dashed lines in all the panels show the completeness limiting magnitude of 18.15 mag, 17.50 mag, and 16.95 mag in $J$, $H$, and $K$ bands, respectively.}
    %\label{fig:IRAS_06063+2040_comp}
%\end{figure}

\subsection{Extent of the cluster}
\label{cluster_rad}
\begin{figure*}
    \centering
    \includegraphics[width=12.5 cm]{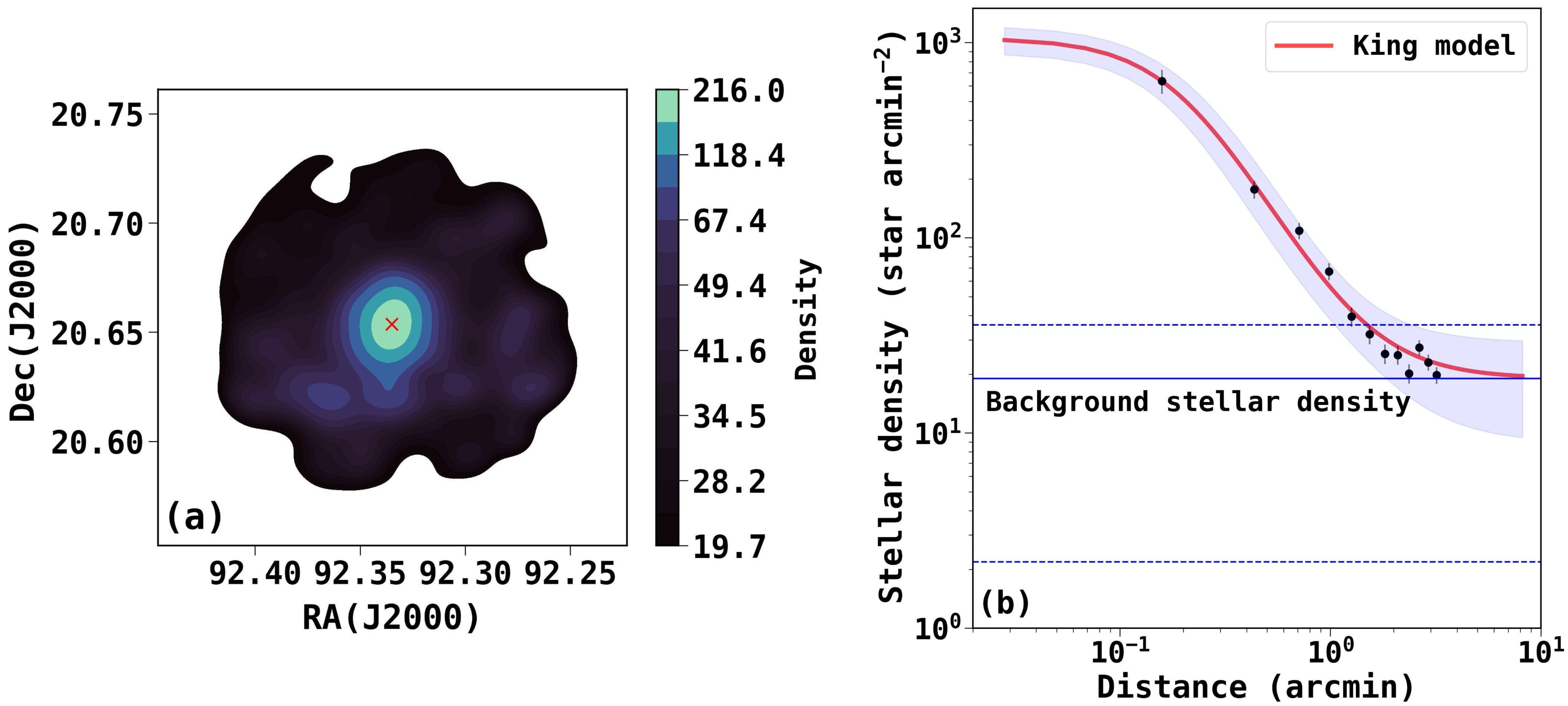}
    \includegraphics[width=5.2 cm]{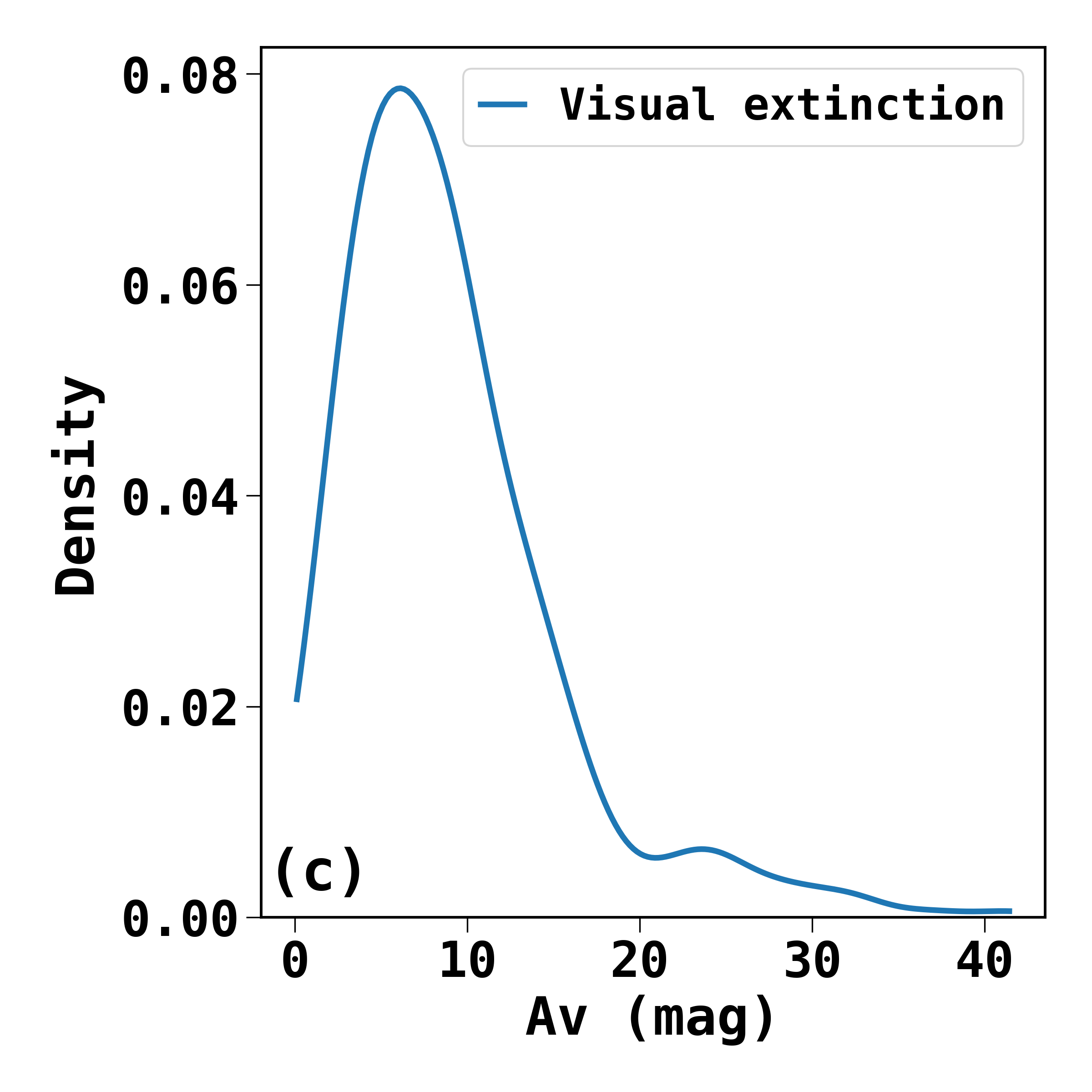}
    \caption{(a) A 2D density plot of the stellar distribution observed in the direction of the IRAS 06063$+$2040 cluster, with the cross symbol indicating the cluster centre taken at the peak density point. (b) The observed stellar surface density of IRAS 06063$+$2040 as a function of distance from the centre (cross symbol in panel-a). The red curve shows the best fit King's profile along with 3$\sigma$ uncertainty as blue shaded region. The error bars at each point represent the Poisson uncertainties. The solid blue line shows the best-fit background stellar density with 5$\sigma$ uncertainty as blue dashed lines. (c) Density plot of visual extinction of all the sources observed towards the direction of IRAS 06063$+$2040.}
    \label{fig:stellar_density_iras2040}
\end{figure*}

In comparison to open clusters, defining the centre of young clusters is much more difficult due to variable high extinction, nebulosity, and complex shapes. %The cluster centre can be taken at the geometrical centre of the region, the location of a massive star, and the highest column density point of the region. 
%Taking the cluster centre at the geometrical centre depends upon the region of the target adopted for the study and hence can be biased. Therefore, for young clusters in the sample, the highest stellar density point was chosen as the centre of the clusters.  
For assessing the shape and size of the cluster, we selected data for a larger region around the geometric centre of the cluster and plotted a 2-dimensional kernel density estimate (KDE) map with a bin width of 0.5 to 0.7. The region for making the KDE map is selected in such a way that there should be an ample number of stars to make a smoothened KDE map of the cluster. The optimum bin width was chosen to have a good compromise between over- and under-smoothing density fluctuations, depending upon the data statistics in the cluster region. Then, we find the peak density point using Gaussian KDE and choose these coordinates as the approximate centre of the cluster. Fig. \ref{fig:stellar_density_iras2040}a shows the 2D density plot of IRAS 06063$+$2040 with its peak density point marked with a cross sign.%, and the corresponding spatial distribution of the stars is shown in Fig. \ref{}, where the clustering of stars can be easily seen. 
 %For clusters IRAS 06104$+$1524A and IRAS 06104$+$1524B, which are very close to each other such that it is difficult to make their separate KDE plots, we made a single KDE plot covering both clusters and determined their centres by finding the two highest stellar density points in the KDE plot (see Figure xx).

To determine the effective radius of the cluster ($R_{\rm{clust}}$), we constructed its radial density profile (RDP). To do this, we first divided the cluster into different annular rings from the centre. The radius of the annular rings from the centre was defined by the auto-binning option of the Pyplot module in Matplotlib. Then, we counted the stars in each radial bin and calculated the stellar density in each annulus by dividing the total counts with the area of the annulus. Fig. \ref{fig:stellar_density_iras2040}b shows the plot of stellar density as a function of radius for IRAS 06063$+$2040. In order to obtain the radius of the cluster, we fitted the RDP with the empirical King's profile \citep{King_1962} of the form

\begin{equation}
    \rho(r) \propto b_0 + \frac{\rho_0}{\left[1 + \left(\frac{r}{r_c}\right)^2\right]}, 
\end{equation}
where $b_0$, $\rho_0$, and $r_{\rm{c}}$ are the background stellar density, peak stellar density, and core radius of the cluster, respectively. The King's profile fit to the stellar density of IRAS 06063$+$2040 is shown in Fig. \ref{fig:stellar_density_iras2040}b, and the background stellar density is shown by a solid blue line. The radius of the cluster is defined as the radial distance at which the modelled stellar density lies 5$\sigma$ above the background stellar density. For example, Fig. \ref{fig:iras_06063+2040_image}a and b also shows the outer extent of 
the IRAS 06063$+$2040 cluster on the UKIDSS and 2MASS images. As can be seen from the images, the stellar density beyond the cluster boundary is very low and indistinguishable from the background density, implying that the chosen cluster radius above 5$\sigma$ of the background is a reasonable choice.

%Here, we acknowledge that the radial density plot depends upon the radial bins. Although we opted for optimum bins, but 
For some clusters, we found that King's profile does not completely become flattened at the background stellar density, either due to low statistics or confusion with other nearby stellar groups/clusters.
%In such cases, apart from taking the radius at 5-$\sigma$ above the background stellar density, we also rechecked and confirmed the size of the clusters from the stellar density contours in the 2-D KDE maps of the clusters and by visually inspecting the UKIDSS NIR images. %Nonetheless, there can be uncertainty in the radius of the clusters, however, that will be mostly within the large uncertainty in the other cluster parameters like gas mass (discussed in Section \ref{clump_gas}).   
In such cases, we choose an optimum value from the 2D stellar density map and also by visually inspecting the UKIDSS NIR images. The radius of all the clusters is given in Table \ref{tab:clumps_gas_prop} and lies in the range of 0.5 to 1.6 pc. 
%As can be seen from the table, all the clusters in our sample are parsec to sub-parsec in size (i.e. $R_{\rm{clust}}$ $\sim$0.5$-$1.6 pc). 
%, implying that
%the faction of cluster members that may be lying in the tail of the stellar density distribution would be low to affect the global properties of the cluster.}

The photometric data can be incomplete towards the fainter end in the clustered environment due to stellar crowding and a bright background. To determine the completeness of the UKIDSS NIR data within the cluster radius, the histogram turnover method was used \citep[for details, see][and references therein]{dam21, Rawat_2024c}. %\citep{ohlen_2013, samal_2015, jose_2016}.
%As discussed in \cite{Rawat_2024c}, 
This approach allows a determination of the approximated completeness limits. %In this method, the histogram distribution of the photometric data is plotted, and where it deviates from the linear behaviour is generally taken approximately as a 90\% completeness limit. 
%Fig. \ref{fig:IRAS_06063+2040_comp} shows an example of the KDE histograms of the sources detected in various bands for IRAS 06063$+$2040. 
With this approach, the completeness limit for most clusters within the cluster radius was found to be in the ranges 17.1$-$18.3, 15.9$-$17.5, and 15.6$-$16.8 mag in the $J$, $H$, and $K$ bands, respectively.

\subsection{Extinction}
\label{extinction}
For deriving cluster properties, the contributions of field stars and extinction in the direction of the cluster need to be determined.
%The field star population in the cluster region along the line-of-sight can significantly contaminate the cluster population and, hence, the derived cluster properties. 
%In the case of embedded clusters, the background contamination is not expected to be significant, as the background stars are highly extincted by the dust in the cloud itself to be detected with the current sensitivity of the point sources \citep[e.g. see discussions in][]{Rawat_2024c}. %\textbf{Since most of the targets in our sample are beyond $\gtrsim$1.5 kpc, we assume the background contamination to be minimal.} 
%Whereas foreground contamination can be very significant and, therefore, is a factor that needs to be removed. 
To assess extinction toward the cluster, following the commonly adopted approach \cite[e.g.][]{Gutermuth_2008, Gutermuth_2009}, a control field region near the cluster location that is relatively dust-free was selected to evaluate the median intrinsic colors of field stars in the cluster's direction.  Large control fields with a radius of around 5$-$7 arcmin were used to get better statistics of the field population. The photometric data of the control fields was selected in the same way as done for the cluster sample. We then estimated the visual extinction ($A_{\rm{V}}$) of the observed stars projected within
the cluster radius using the following equation and using the mean $J-H$ and $H-K$ colors of the control field sources as the intrinsic colors of the stars:

\begin{equation}
\av = c\, \times [(i-j) - (i-j)_0];\quad i\, \mathrm{and}\, j = J, H,\, \mathrm{and}\, K, 
\end{equation}
where the mean color of the control field sources $(i-j)_0$ is subtracted from the observed colors of the stars $(i-j)$. Here, $c$ is the constant whose value is taken to be 9.34 and 15.98 for the $J-H$ and $H-K_s$ color excesses, respectively, by adopting the \cite{Rieke_1985} extinction laws. We then derived the median extinction of the cluster
following the method adopted in \citet{Rawat_2024c}.
%hereafter target population/sources, we assumed the $J - H$ and $H - K_s$ colors of the control field to be intrinsic, as they are relatively dust-free, and subtracted them from the observed colors of the target population. We used the equation \ref{} of Chapter \ref{chap_FSR} to find the visual extinction ($\rm{A_V}$) values of each observed star along the cluster direction within its region.
Fig. \ref{fig:stellar_density_iras2040}c shows the density plot of visual extinction for IRAS 06063$+$2040. %It shows a peak around xx magnitude and a long tail at high extinction end. 
We find 
that the median \av for IRAS 06063$+$2040 is around 6.0 $\pm$ 3.1 mag, with only 6\% sources having \av greater than 20 mag. Following a similar analysis, we find the median \av values for the cluster sample to be in the range of 2 to 11 mag 
and are given in Table \ref{tab:clumps_gas_prop}.

%\begin{figure}
    %\centering
    %\includegraphics[width=8.5 cm]{Av_Distribution_cf2.png}
    %\caption{Density plot of visual extinction of all the sources observed towards the direction of IRAS 06063$+$2040.}
    %\label{fig:Av_iras_2040}
%\end{figure}

\subsection{\emph{K}-band luminosity function and age estimation}
\label{KLF_age}
\begin{figure*}
    \centering
    \includegraphics[width=16 cm]{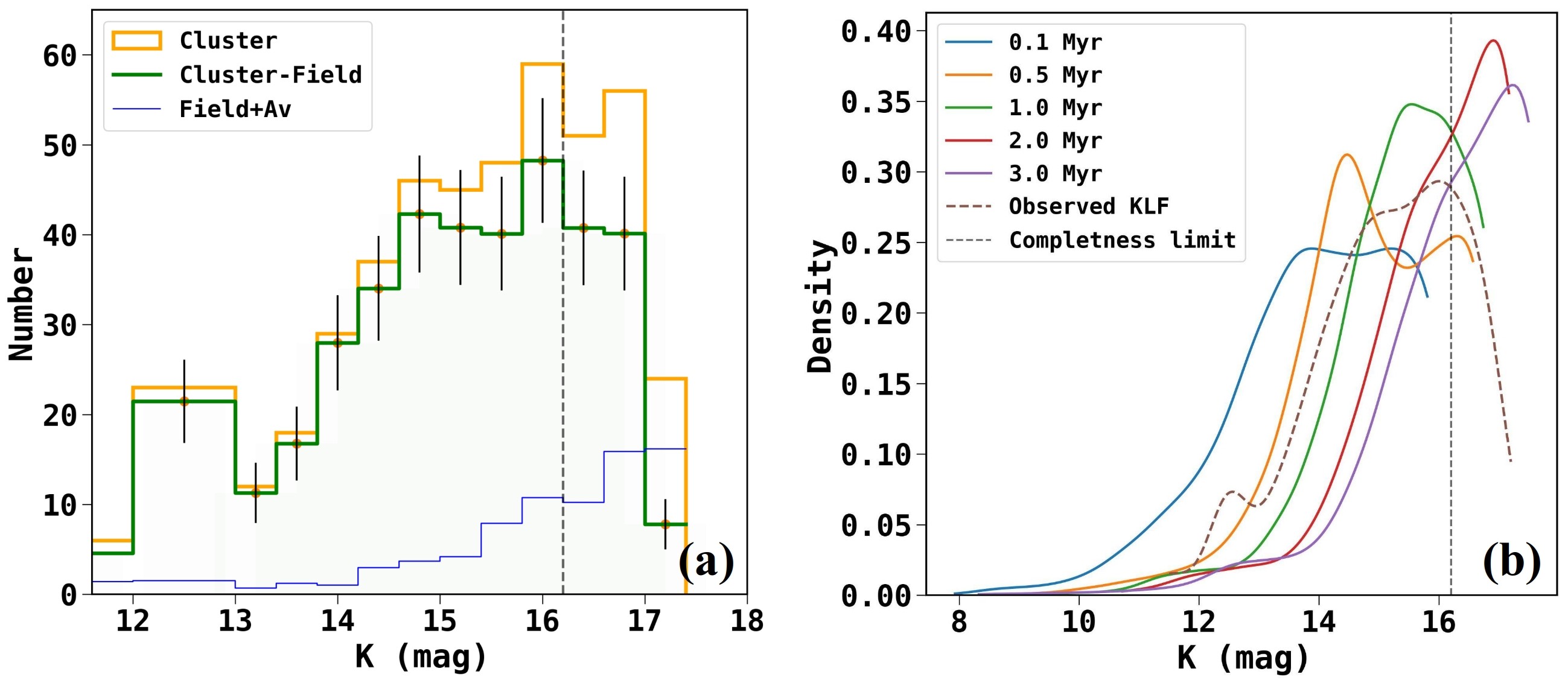}
    \caption{(a) $K$-band luminosity function of the IRAS 06063$+$2040 cluster (orange), reddened control field (blue), and control field subtracted cluster (green). (b) $K$-band density plots of synthetic clusters of age 0.1, 0.5, 1.0, 2.0, and 3.0 Myr, shown by solid curves. The dashed curve shows the control field subtracted $K$-band density plot of IRAS 06063$+$2040. The dashed grey line shows the completeness limit of the $K$-band data within the cluster region.}
    \label{fig:KLFs_iras2040}
\end{figure*}

The $K$-band luminosity function (KLF) is the distribution of $K$ band magnitudes of the sources, which can be used to estimate the proxy age of a cluster by matching the KLF of the observed cluster with that of the modelled clusters \citep[for details, see][and references therein]{Rawat_2024c}. Briefly, we adopt the following procedure.  The KLF is expressed as:

\begin{equation}
    \frac{dN}{dm_K} = \frac{dN}{dM_*} \times \frac{dM_*}{dm_K},
\end{equation}
where $m_K$ and $M_*$ are the $K$-band luminosity and mass of the stars, respectively. In the above equation, $dN/dm_K$ represents the number of stars within a $K$-band magnitude bin, $dN/dM_*$ corresponds to the underlying stellar mass function, and $dM_*/dm_K$ describes the mass-luminosity ($M-L$) relation. Thus, to determine the  KLF of the cluster, firstly, the contamination of the field stars was removed by subtracting the KLF of the control field population from that of the observed cluster population. We use the same control field discussed in Section \ref{extinction} to assess the field star population in the direction
of the cluster.
The KLF of the control field is the distribution of the $K$-band magnitudes of the field stars reddened by the median \av of the cluster region. Fig. \ref{fig:KLFs_iras2040}a shows the KLF of the cluster before field subtraction and the KLF of the reddened control field with the same bin size. Since the size of the control field region is larger than the size of the cluster region, the field sources are first normalized to the cluster size at each bin and then subtracted from the cluster KLF to obtain the field-subtracted cluster KLF, which is also shown in Fig. \ref{fig:KLFs_iras2040}a. %By reddening the field population, the background contamination will also further go down beyond the sensitivity limits of the UKIDSS observations. Consequently, most of the stars within the cluster region are likely to be cluster members after field subtraction.        

Then, for estimating the age ($t_{\rm{clust}}$) of a cluster, we compared the field-subtracted cluster KLF with the modelled KLFs of synthetic clusters at different ages. We used the Stellar Population Interface for Stellar Evolution and Atmospheres (SPISEA) python code \citep{Hosek_2020} to generate the synthetic clusters with an age range of 0.1 to 3.0 Myr. The SPISEA code generates the synthetic clusters based on various input parameters such as initial cluster mass, distance, functional form of the IMF, metallicity, stellar evolution models, atmosphere models, and extinction law \citep[for details, see][]{Hosek_2020}. The code randomly draws the stars from the chosen IMF until the total mass of stars reaches close to the initial cluster mass. We generated synthetic clusters of solar metallicity with photometric magnitudes in the UKIDSS $J$, $H$, and $K$ filters, at the adopted distances of the observed clusters, using the MIST evolutionary models, a composite grid of atmospheric models, the Kroupa IMF, and the Rieke–Lebofsky extinction law \citep{Rieke_1985}. The code generates all the stars at the same age, depending on the selected isochrone model.  %We generated the synthetic clusters with an age range of 0.1 to 3 Myr. 
Since SPISEA randomly generates sources for a cluster of a given age, to obtain the median KLF of each age, we ran the simulation 200 times for each age. We then matched the observed KLF of each cluster with the synthetic KLFs of different ages, as discussed in \cite[][]{Rawat_2024c}. %, and it also does not account for any dynamical evolution of the cluster, so no accelerated star formation is considered. Therefore, there is no age spread in the synthetic clusters. Here, we have studied the compact clusters, so a large age spread is not expected within the cluster radii, such that the effect of age spread would not be significant in this work. %Furthermore, we have adopted a 50\% uncertainty in age, which has been propagated to all derived properties.
  %We did not consider the stellar multiplicity factor to avoid complexity due to multiple stellar systems, as they are assumed to be unresolved in the current SPISEA model.

Fig. \ref{fig:KLFs_iras2040}b shows the KLFs of the synthetic clusters and the field subtracted KLF of IRAS 06063$+$2040. From the figure, it can be seen that the KLF of IRAS 06063$+$2040 lies close to the synthetic KLFs of age between 0.5 to 1.0 Myr. Therefore, we considered  0.75 Myr as the approximate age of the IRAS 06063$+$2040 cluster. 
Star clusters can exhibit an age spread, which is not accounted for in the SPISEA model. It is suggested that high-density molecular clumps may yield clusters with smaller stellar age spreads than those formed by low-density clumps, and observations also support this hypothesis that compact clusters tend to exhibit less age spread compared to extended cluster-forming regions \citep[][and references therein]{par14}. This could be the reason that the observed KLF in Fig. \ref{fig:KLFs_iras2040}b is distributed mainly between the 0.5 to 1 Myr KLF of the synthetic clusters. We thus assigned a median age of 0.75 $\pm$ 0.25 Myr to IRAS 06063$+$2040, which also agrees with the age and age dispersion of the cluster obtained through an independent method, which is described below. Following a similar analysis, we determined the ages of all the clusters, which are given in Table \ref{tab:clumps_gas_prop}. The ages of most of the clusters range from 0.5 to 1.5 Myr, suggesting that they are indeed young clusters. We find that for the obtained age and median extinction of the clusters, the $K$-band data is largely complete down to $\sim$0.2 \Ms.

The age of young clusters is a fundamental parameter that is among the most uncertain and difficult to constrain \citep[see][for a review] {Soderblom_2014}. The classic method for estimating the age is through the Hertzsprung–Russell diagram (HRD), where the positions of member stars are compared with the locations of theoretical pre-main sequence evolutionary tracks and isochrones. For embedded clusters, due to variable extinction \cite[e.g.][]{getman19}, this method may work well if spectroscopic measurements are available 
for a significant number of stars. While for optically visible clusters with insignificant extinction
variation, optical color-magnitude diagrams are often used to constrain age \cite[e.g.][]{pandey13,jose17,panwar17}.  In the present work,
due to the lack of spectroscopic information for the studied sample, we constrained the age of the clusters using KLF modelling.
Although KLF modelling is a proxy way for estimating the age, its advantage for embedded clusters is that it accounts for embedded sources in the age determination. For embedded clusters, using an optical HR diagram may bias the estimated cluster age toward older values, as young clusters and star-forming regions often contain stars at various evolutionary stages—ranging from evolved Class III sources to embedded Class I sources \cite[e.g.][]{allen08,samal15,povich16}. Nonetheless, as a sanity check, we determined the age of IRAS 06063$+$2040 by deriving the stellar properties of the sources through spectral energy distribution (SED) modelling, where we fitted the observed SED with theoretical models to derive the stellar properties and subsequently derived the age of the stars. SED modelling with data points in optical bands better constrains the stellar properties. Therefore, we restrict the SED modelling to sources within the cluster radius that have measurements available in the optical bands, in addition to the infrared (1.0$-$8 $\mu$m) bands. Doing so, we estimated the 
median age of the cluster, based on 15 well-fitted sources, to be $\sim$0.8 $\pm$ 0.4 Myr, which agrees well with the age derived from the KLF modelling. This close agreement between the methods may be accidental for IRAS 06063$+$2040, but this analysis suggests that, in the absence of better information on embedded sources, KLF modelling is a reasonable approach for deriving cluster age. 
%Thus, we assume that the age estimation done of the studied clusters through KLF modelling is likely to be accurate within 0.5 Myr.
The details of the SED modelling and age determination for IRAS 06063$+$2040 are given in Appendix \ref{SED}.

Moreover, we want to emphasize that the age determination of a MC complex is also scale-dependent, as star formation models of collapsing clouds such as the global hierarchical collapse \citep[GHC;][]{sema2019} suggest that clouds collapse hierarchically from large-scale to small-scale, forming structures (such as filaments, clumps, and cores) under the influence of global gravity. %And each of these structures form stars with their own free-fall time, while due to global gravity the rich massive clusters form at the junction of the filamentary structures, called hubs.  
These models suggest that global gravity results in ever-increasing densities towards the cloud's centre of
potential via filamentary flows, resulting in older stars in the extended part of the cloud and young massive stars/clusters in the hub, where filaments merge to form dense and massive clumps \citep[e.g., see discussion in][for observational evidence]{Rawat_2023, Rawat_2024a}. In fact, based on photometric analysis, such a core-halo age gradient \cite[e.g.][]{getman18,panwar18} or evolutionary gradient \cite[e.g.][]{Rawat_2023} has been observed in young clusters and molecular clouds, i.e. stars in the core regions are younger than those in the halo or extended regions. All these results imply that the average age of a molecular cloud may differ from the age of the individual cluster-forming clumps. Thus, it is necessary to determine the age of the region of interest for a better estimation of the SFR and SFE. 

 Based on the error in age estimation for IRAS 06063+2040 (i.e. $\sim$0.8 $\pm$ 0.4 Myr) from the SED fitting and the distribution in KLFs of the clusters, in this work, we have adopted an uncertainty of 50\% in the age estimation of all the clusters. We note that the ages reported in Table \ref{tab:clusters} are the present average ages of the clusters within their respective radii. 

\subsection{Gas properties of the clusters}
\label{clump_gas}
 
One important term in the scaling relations is the gas mass of the star-forming regions.  %The PPMAP technique provides a $\sim$12 arcsec resolution column density map, as it retains the psf information at each wavelength, and also considers the variation in temperature along the line-of-sight (see chapter \ref{chap_dust} for details). 
We calculated the gas mass ($M_{\rm{gas}}$) using the {\it Herschel} column density map and the following equation:

\begin{equation}
    M_{\rm{gas}} = \mu_{\rm{H_2}} m_{\rm{H}} A_{\rm{pixel}} \Sigma N(H_2),
\end{equation}
where $\mu_{\rm{H_2}}$ is the mean molecular weight of the hydrogen molecule that is taken to be 2.8 \citep{Kauffmann_2008}, $m_{\rm{H}}$ is the mass of the hydrogen atom, $A_{\rm{pixel}}$ is the area of the pixel, and $\Sigma N(H_2)$ is the integrated column density. We estimated $\Sigma N(H_2)$ within the clump boundary 
using PPMAP-based column density maps. Defining the clump boundary in a cloud is a complex problem, as it depends on
parameters such as the resolution, tracer, and sensitivity of the observations. High-resolution data leads to more resolved structures, while low-resolution data can lead to massive
structures. Similarly, warm tracers generally trace more extended environments, whereas millimetre wavelengths are better suited for tracing cold gas, which is most relevant to star formation. 
In the present case, since most clusters are of size $>$ 0.5 pc; thus, we search for cold clumps identified with low-resolution (beam $\sim$33\arcsec) 1.1-millimetre data from
the Bolocam Galactic Plane Survey \citep[BGPS;][]{Dunham_2011}, around the stellar density peak. We found that for six sources, the radii from the BGPS clumps are available and are found to be
in close agreement with the corresponding cluster radii. We thus estimate the gas mass of the cluster within the cluster radius.
%in close agreement with the radii of the clusters (out of 8 available, 6 have a radius within 25\%). Thus, we opted...}.
The gas masses of the clumps range from 120 to 1400 \Ms~and are given in Table \ref{tab:clumps_gas_prop}. The uncertainty associated with the gas mass of the clumps ranges from 39\% to 55\%, accounting for uncertainties in the gas-to-dust mass ratio, opacity index, and distance of the clumps \citep[for details, see][]{Rawat_2023}. Using $M_{\rm{gas}}$ and $R_{\rm{clust}}$ of the clumps, the molecular gas mass surface density ($\Sigma_{\rm{gas}}$), number density ($n_{\rm{H_2}}$), and free-fall time of the clumps are estimated by using the following equations:

\begin{equation}
    \Sigma_{\rm{gas}} = \frac{M_{\rm{gas}}}{\pi R^2_{\rm{clust}}},
\end{equation}
\begin{equation}
    n_{\rm{H_2}} = \frac{3M_{\rm{gas}}}{4\pi R^3_{\rm{clust}}\, \mu_{\rm{H_2}} m_{\rm{H}}},
\end{equation}
\begin{equation}
    t_{\rm{ff}} = \left( \frac{3~\pi}{32~ G \mu_{\rm{H_2}} m_{\rm{H}} n_{\rm{H_2}}} \right)^{1/2},
\end{equation}
where $G$ is the gravitational constant. %Here, it is important to note that the densities ($\Sigma_{\rm{gas}}$ and $n_{\rm{H_2}}$) are of molecular gas, whereas the KS relation (equation \ref{eq:KS}) is for total (atomic plus molecular) gas density. %\textbf{However, there are many ``KS'' relations built only based on the molecular component \citep[e.g.][]{Bigiel, Reyes_2019}.} %However, \cite{Reyes_2019} have found that the correlation between SFR and gas mass is stronger and linear with molecular hydrogen compared to atomic hydrogen.}
All the physical parameters of the clumps are listed in Table \ref{tab:clumps_gas_prop}. We want to point out that for the IRAS 05480$+$2545 cluster in our sample, the PPMAP-based map shows saturation at high-density regions. Therefore, we used the Hi-GAL column density map from \cite{Schisano_2020} (resolution $\sim$34\arcsec) for this particular cluster. %that have a relatively low resolution of $\sim$34\arcs. 
We compared the mass estimates derived from the column density maps of PPMAP and \cite{Schisano_2020} for cluster FSR 655 \citep{Rawat_2024a}, shown here in Fig. \ref{fig:SFR-gas_mass} and \ref{fig:SFR-gas_mass_tff}. By doing so, we found that the estimated total gas masses from both column density maps are comparable within 20\%. %This difference in total mass is for the whole cloud, whereas, in the case of clumps having a size of only a few parsecs, the difference in clump properties from \cite{Marsh_2019}'s PPMAP and \cite{Schisano_2020}'s column density map would be even lesser and within the uncertainty limits. %For example, in the case of IRAS 05480+2545, we found a difference of less than 5 per cent in the clump mass of common regions from both the column density maps.       

%\begin{sidewaystable}
\begin{table*}
 \centering
 \caption{Clump physical properties.}
\begin{tabular}{|p{0.5cm} p{2.5cm} p{1.5cm} p{0.8cm}  p{2.0cm} p{1.7cm} p{1.7cm} p{1.7cm} p{1.5cm}|} 
\hline
\hline

No & Name & $R_{\rm{clust}}$ & \av & $t_{\rm{clust}}$ & $M_{\rm{gas}}$ & $\Sigma_{\rm{gas}}$ & $n_{\rm H_2}$ & $t_{\rm{ff}}$\\ 
&     & (pc) & (mag) & (Myr) & (\Ms) & (\Ms~pc$^{-2}$) & (cm$^{-3}$) & (Myr)\\
\hline
1 & IRAS 06063+2040 & 0.97 $\pm$ 0.19 & 6.0 & 0.750 $\pm$ 0.375 & 460 $\pm$ 253 & 155 $\pm$ 105 & 1732 $\pm$ 1403 & 0.74 $\pm$ 0.30\\ 
2 & IRAS 06055+2039 & 0.79 $\pm$ 0.16 & 8.6 & 0.750 $\pm$ 0.375 & 810 $\pm$ 446 & 410 $\pm$ 279 & 5628 $\pm$ 4559  & 0.41 $\pm$ 0.17\\
3 & IRAS 06068+2030 & 0.87 $\pm$ 0.17 & 5.1 & 0.50 $\pm$ 0.25 & 300 $\pm$ 165 & 126 $\pm$ 86 & 1564 $\pm$ 1267 & 0.78 $\pm$ 0.32\\
4 & IRAS 22134+5834 & 0.48 $\pm$ 0.10 & 7.5 & 1.50 $\pm$ 0.75 & 120 $\pm$ 66 & 167 $\pm$ 114 & 3796 $\pm$ 3075 & 0.50 $\pm$ 0.20\\
%5 & AFGL 5142 &  &  & 1.6 & yy\\
5 & IRAS 06056+2131 & 0.51 $\pm$ 0.10 & 8.5 & 0.50 $\pm$ 0.25 & 400 $\pm$ 220  & 485 $\pm$ 330 & 10263 $\pm$ 8313 & 0.30 $\pm$ 0.12\\
6 & Sh2-255 IR & 1.00 $\pm$ 0.06 & 10 & 1.0 $\pm$ 0.5 & 1400 $\pm$ 560 & 398 $\pm$ 167 & 4151 $\pm$ 1826 & 0.48 $\pm$ 0.11\\
7 & [IBP2002] CC14 & 0.57 $\pm$ 0.11 & 6.6 & 0.750 $\pm$ 0.375 & 210 $\pm$ 116 & 206 $\pm$ 140 & 3914 $\pm$ 3170 & 0.49 $\pm$ 0.20\\
8 & IRAS 06058+2138 & 0.62 $\pm$ 0.12 & 10.1 & 0.750 $\pm$ 0.375 & 580 $\pm$ 319 & 487 $\pm$ 331 & 8606 $\pm$ 6971 & 0.33 $\pm$ 0.13\\
9 & IRAS 06065+2124 & 0.64 $\pm$ 0.13 & 5.0 & 1.50 $\pm$ 0.75 & 150 $\pm$ 82 & 113 $\pm$ 77 & 1908 $\pm$ 1546 & 0.70 $\pm$ 0.29\\
10 & NGC 2282  & 0.84 $\pm$ 0.20 & 2.3 & 2.0 $\pm$ 1.0 & 170 $\pm$ 102 & 77 $\pm$ 58 & 996 $\pm$ 922 & 0.98 $\pm$ 0.49\\
11 & IRAS 05490+2658 & 1.02 $\pm$ 0.20 & 10.7 & 0.50 $\pm$ 0.25 & 930 $\pm$ 512 & 284 $\pm$ 193 & 3036 $\pm$ 2459 & 0.56 $\pm$ 0.23\\
12 & BFS 56 & 0.90 $\pm$ 0.18 & 8.4 & 1.0 $\pm$ 0.5 & 520 $\pm$ 286 & 215 $\pm$ 146 & 2670 $\pm$ 2163 & 0.60 $\pm$ 0.24\\
%13 & [BDS2003] 89 & 0.41 & 6.8 & 68 $\pm$ 26 & 128 $\pm$ 49 & 3398 $\pm$ 1291 & 0.53 $\pm$ 0.10\\
13 & Sh2-88 & 0.54 $\pm$ 0.02 & 8.5 & 0.30 $\pm$ 0.15 & 590 $\pm$ 230 & 653 $\pm$ 261 & 13266 $\pm$ 5413 & 0.27 $\pm$ 0.05\\
14 & IRAS 06103+1523 & 0.46 $\pm$ 0.09 & 9.1 & 1.250 $\pm$ 0.625 & 450 $\pm$ 248 & 676 $\pm$ 460 & 30778 $\pm$ 24930 & 0.24 $\pm$ 0.10\\
15 & IRAS 06104+1524A & 0.46 $\pm$ 0.09 & 7.3 & 1.0 $\pm$ 0.5 & 170 $\pm$ 94 & 258 $\pm$ 175 & 6095 $\pm$ 4937 & 0.39 $\pm$ 0.16\\
16 & IRAS 06117+1901 & 1.60 $\pm$ 0.32 & 3.1 & 1.50 $\pm$ 0.75 & 640 $\pm$ 352 & 80 $\pm$ 54 & 545 $\pm$ 441 & 1.32 $\pm$ 0.51\\
17 & IRAS 05480+2545 & 0.80 $\pm$ 0.16  & 6.4 & 0.750 $\pm$ 0.375 & 1100 $\pm$ 605 & 551 $\pm$ 375 & 7432 $\pm$ 6020 & 0.36 $\pm$ 0.14\\
%18 & IRAS 06104+1524B & & 2.0 & yy\\

\hline
\hline

\end{tabular}
\label{tab:clumps_gas_prop}
\end{table*}
%\end{sidewaystable}

\subsection{Cluster mass, star formation rate, and efficiency}
\label{sfe}
\begin{figure}
    \centering
    \includegraphics[width= 8 cm]{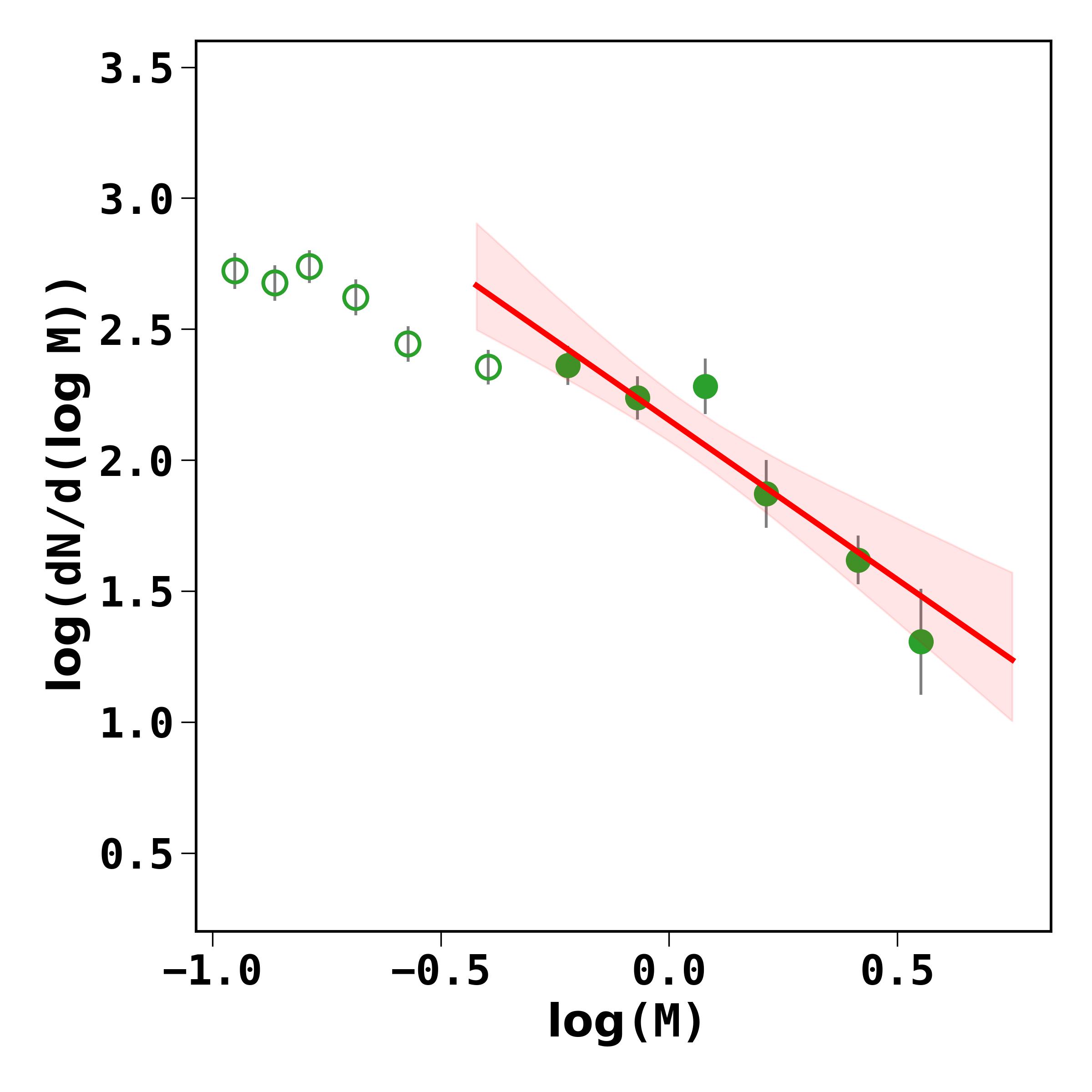}
    \caption{The cluster mass distribution function of IRAS 06063$+$2040 in which the error bars represent the $\pm \sqrt{N}$ errors. The filled circles show the data points used for the least square ﬁt with a power-law function, and the best-fit index, $\Gamma$, is $\sim$1.22 $\pm$ 0.16. The shaded region shows the 3$\sigma$ uncertainty associated with the fitted index.}
    \label{fig_iras_2040_imf}
\end{figure}

In this work, we have determined the age of individual clusters using their $K$-band luminosity functions.  The theoretical mass-luminosity ($dM_*/dm_K$) relation corresponding to the age of a cluster can be matched with its field-subtracted KLF ($K$-band magnitude distribution), to obtain the stellar mass distribution of the cluster. Then, the total stellar mass of the cluster can be estimated by integrating its mass distribution function \citep[e.g. see][]{Rawat_2024c}.
%within certain mass limits. %To obtain the stellar mass distribution of a cluster, we first matched the with the  
We used the mass-luminosity relations from the MIST stellar evolutionary models of solar metallicity \citep{Choi_2016}, and extinction laws of \cite{Rieke_1985} to get the mass distribution corresponding to the KLF. Generally, 
a mass distribution function is described in the following logarithmic form:  %Then, we fitted the mass distribution of clusters with the Kroupa IMF \citep{Kroupa_2001} of the logarithmic form.

\begin{equation}
    \frac{dN(\log M)}{d \log M} \propto M^{-\Gamma}.
\end{equation}
Although our data is complete above 0.2 \Ms~in most cases, to avoid any possible bias that may be introduced at the fainter mass end, we fitted the mass distribution
at the high mass end, i.e. mass $>$ 0.4~\Ms~ to obtain $\Gamma$ values. 
%The data was fit within the 
%mass limit of 0.4 \Ms~to 5$-$7 \Ms, as applicable for each cluster. %, and the best-fit $\alpha$ value along with the uncertainty for all the clusters are given in Table \ref{tab:clusters:out}. 
The $\Gamma$ values of all the clusters are found to be within 2$\sigma$ error of the canonical value of $\Gamma$, i.e. 1.3 \citep{Kroupa_2001} for the mass range of 0.4 \Ms~$<$ $M$ $<$ 8 \Ms. The $\Gamma$ value of 1.3 corresponds to the Kroupa index, $\alpha$ = 2.3 in the linear form of Kroupa IMF \citep{Kroupa_2001}. For example, the mass distribution plot for IRAS 06063$+$2040 is shown in Fig. \ref{fig_iras_2040_imf}, which is fitted with a power-law of best-fit index, $\Gamma$ $\sim$1.22 $\pm$ 0.16. %The large uncertainties in the best-fit index values are because of the low statistics of stars in clusters. 
Since the best-fitted IMF slopes are within the uncertainty of the Kroupa slope, so, we used the Kroupa broken power-law to calculate the total stellar mass of all the clusters. %This will ensure that any under / overestimation of the IMF slope due to various factors (like NIR excess, variable extinction, low statistics of cluster members, contamination at the low-mass end, and age uncertainty) do not affect the cluster properties such that they can be better compared with each other. 
%Although, from the MIST isochrones of the corresponding ages of the cluster (discussed in Section \ref{KLF_age}), the NIR data for most of the clusters was found to be complete down to 0.1$-$0.2 \Ms. Nevertheless, to avoid any contamination at the low-mass end, we have taken an upper limit of 0.5 \Ms~in the IMF. Thus, 
Briefly, we first integrated the IMFs of the clusters with the Kroupa index for mass limits of 0.5 to 15 \Ms~and then extrapolated it down to 0.1 \Ms~using the functional form of the Kroupa mass function. 
Doing so, we obtained the stellar masses of the clusters in the range between 43 to 500 \Ms~and these are listed in Table \ref{tab:clusters:out}. %along with their associated uncertainties. 
%We have also estimated the stellar mass by integrating the IMF directly up to the lower mass limit of 0.2 \Ms~(without extrapolating) for those clusters that have data completeness down to these limits. By doing so, we found that the total stellar mass in both the approaches only differs by 5$-$10\%.%The uncertainty in the total stellar mass of each cluster is due to the error in the best-fit IMF slope that has been taken as a typical uncertainty in the Kroupa slope for estimating the total stellar mass. 

\begin{figure}
    \centering
    \includegraphics[width=8.55 cm]{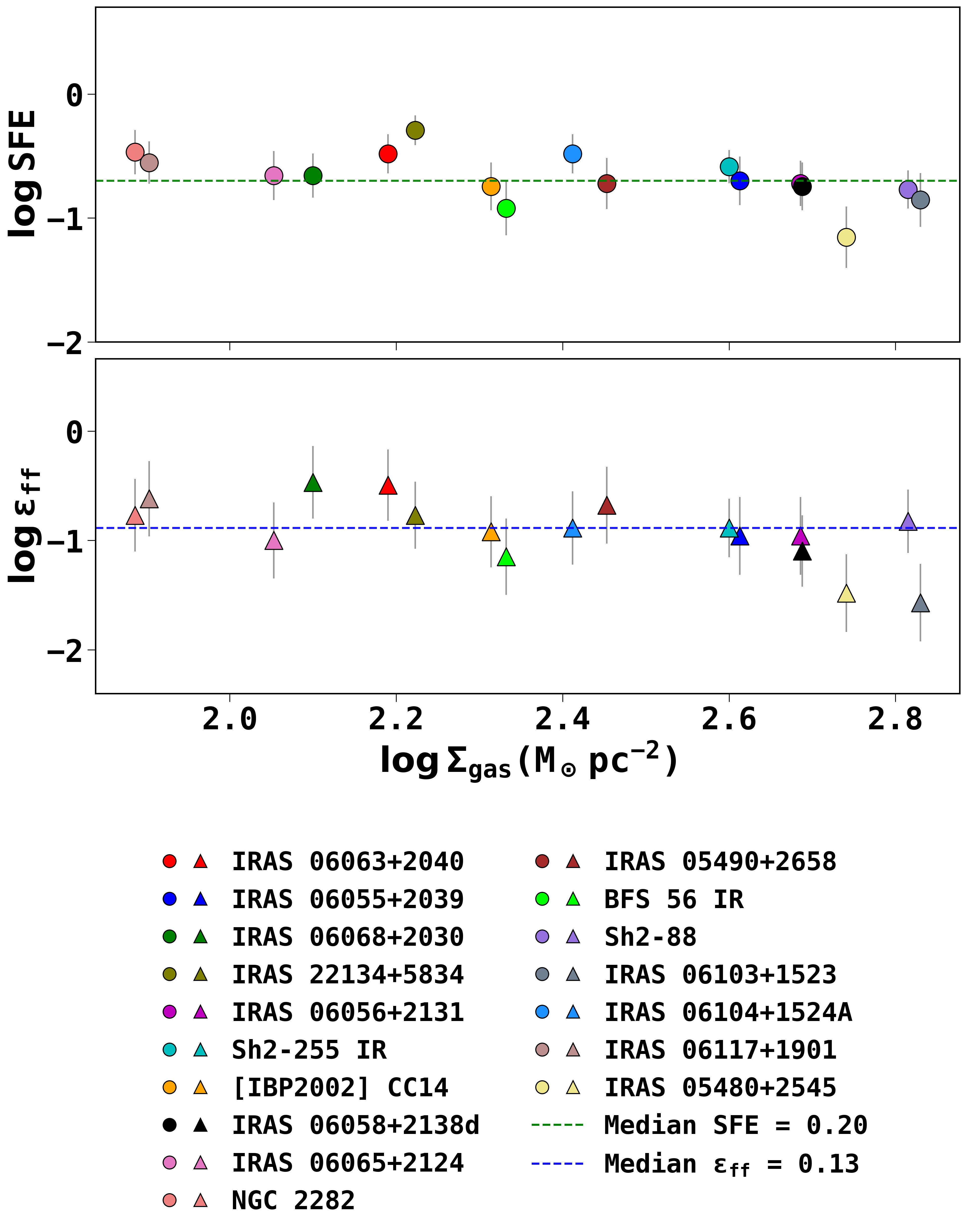}
    \caption{The SFE and $\epsilon_{\rm{ff}}$ of the clusters plotted with their gas mass surface densities. %The vertical dotted lines connect symbols corresponding to the same region. 
    %The coloured encircled dots represent those clusters which have $\frac{F_{\rm{grav}}}{F_{\rm{rad}}}$ $<$ 3 (see Section \ref{high_eff}).
    }
    \label{fig:sfe_clusters}
\end{figure}

Using $M_{\rm{gas}}$, $M_*$, and $t_{\rm{clust}}$, we estimated the instantaneous SFR and SFE of the clusters. 
We find that the SFEs of the clusters lie in the range of 0.07 to 0.51, with a mean and median around 0.23 and 0.20, respectively. The SFRs lie in the range 29 to 500 \Ms~Myr$^{-1}$, with a mean and median around 184 \Ms~Myr$^{-1}$ and 163 \Ms~Myr$^{-1}$, respectively. 
%The SFR varies from clump to clump depending upon the stellar mass and age of the cluster within them, and the value lies in the range of 29 to 500 \Ms~Myr$^{-1}$ with a median around 166 \Ms~Myr$^{-1}$. %\textbf{The SFRs estimated here are within an order of magnitude of the values expected from the SFR–clump mass relation given in \cite{Elia_2022}.} 
%The SFEs of the clusters range from 0.07 to 0.51, with a mean and median around 0.23 and 0.22, respectively.  %However, for clusters in the sample with ages less than 2 Myr, the mean SFE turns out to be around 0.23.  %Notably, [BDS2003] 89 has the maximum SFE (i.e. 0.62), which means that the cluster has converted most of its gas into stars as it is relatively old ($\sim$3 Myr). 
The spread in the SFEs may occur due to the different evolutionary stages of the clusters \cite[e.g.][]{Dib_2011, Lee_2016}. The SFE of a region initially rises with time as a result of star formation and later can increase due to the decrease in the cluster's gas mass caused by the conversion of gas into stars and/or the dispersal of gas from feedback mechanisms. The latter effect could bias the instantaneous SFE due to the way it is defined. 
%Hence, the instantaneous SFE initially underestimates and later overestimates the total fraction of gas converted into stars over the lifetime of a star-forming region \citep{Megeath_2022}. 
Therefore, the instantaneous SFE is only reliable for young star-forming regions, which are at the early stages and have significant gas mass left to form stars. In our case, most of the clusters are very young and associated with gas of high surface density (gas with corresponding \av $>$ 7$-$8 mag or $\Sigma_{\rm{gas}}$ = 110$-$130 \Ms~pc$^{-2}$, see Table \ref{tab:clumps_gas_prop}). In molecular clouds, structures with \av $>$ 7$-$8 mag are, in general, found to be the preferred sites of active star formation \citep{Lada_2010, Andre2017, Patra_2022}. Moreover, we find that stellar feedback is not yet significant in these clumps to significantly affect star formation properties (discussed in Section \ref{high_eff}). Therefore, in our case, the estimated SFE and SFR are likely free from the biases discussed above. The efficiency can also be expressed as the fraction of gas mass converted into stars per free-fall time, which is defined as $\epsilon_{\rm{ff}}$ = SFE $\times \, t_{\rm{ff}}/t_{\rm{clust}}$. The $\epsilon_{\rm{ff}}$ of the clusters ranges from 0.03 to 0.34, with a mean and median around 0.15 and 0.13, respectively. The variation of SFE and $\epsilon_{\rm{ff}}$ of the clusters is shown in Fig. \ref{fig:sfe_clusters} with their $\Sigma_{\rm{gas}}$. Sh2-255 IR cluster has the maximum SFR (i.e. $\sim$500 \Ms~Myr$^{-1}$), while IRAS 06065$+$2124 has the minimum SFR (i.e. $\sim$29 \Ms~Myr$^{-1}$). However, we want to emphasize that the measured SFE and SFR are instantaneous observational values. If the clumps are part of dynamically evolving clouds, these instantaneous values are likely biased toward high SFEs and SFRs, as the extended reservoir that could supply additional material to the clump is not accounted for. The measured star formation properties  are given in Table \ref{tab:clumps_gas_prop} and \ref{tab:clusters:out} 
 and the uncertainties in the estimated parameters are derived by propagating errors associated with the gas mass, stellar mass, distance, and age of the clumps.
 %, which have been systematically propagated into the final parameter estimates.

%The mean SFE of the clumps matches well with the SFE xxxxxxxxxx xxx.    

\begin{table}
%\begin{table*}

 \centering
\begin{sideways}
 
\begin{tabular}{|p{0.5cm} p{2.5cm} p{1.4cm} p{1.7cm} p{1.6cm} p{1.6cm} p{1.7cm} p{2.1cm} p{2.1cm} p{1.0cm}|} 
\hline
\hline

No & Name  &  $M_{*}$ & SFE & SFR & $\Sigma_{\rm{SFR}}$ & $\Sigma_{\rm{gas}}$/$t_{\rm{ff}}$ & $S_{\rm{\nu}}$  & log$\left(\frac{N_{\rm{lyc}}}{ \rm{s^{-1}}}\right)$ & $\frac{F_{\rm{grav}}}{F_{\rm{rad}}}$ \\
 &    &  (\Ms) & & (\Ms~Myr$^{-1}$) & (\Ms~Myr$^{-1}$ pc$^{-2}$) & (\Ms~Myr$^{-1}$ pc$^{-2}$) & (mJy) &  & \\
\hline
1 & IRAS 06063+2040   & 223 $\pm$ 24  & 0.33 $\pm$ 0.12 & 297 $\pm$ 152 & 101 $\pm$ 65 & 210 $\pm$ 166 & 41.09 $\pm$ 0.69 & 46.191 $\pm$ 0.009 & 5.5 \\ 
2 & IRAS 06055+2039   & 201 $\pm$ 22 & 0.20 $\pm$ 0.09 & 268 $\pm$ 137 & 137 $\pm$ 89 & 1000 $\pm$ 791 & 119.76 $\pm$ 0.35 & 46.656 $\pm$ 0.006 & 16.2\\
3 & IRAS 06068+2030   & 83 $\pm$ 7 & 0.22 $\pm$ 0.09 & 166 $\pm$ 84 & 70 $\pm$ 45 & 162 $\pm$ 128 & 76.37 $\pm$ 0.33 & 46.460 $\pm$ 0.006 & 2.2 \\
4 & IRAS 22134+5834   & 125 $\pm$ 10 & 0.51 $\pm$ 0.14 & 83 $\pm$ 42 & 115 $\pm$ 76 & 334 $\pm$ 264 & 3.24 $\pm$ 0.20 & 44.794 $\pm$ 0.027 & 5.4 \\
5 & IRAS 06056+2131   & 91 $\pm$ 7 & 0.19 $\pm$ 0.08  & 182 $\pm$ 92 & 223 $\pm$ 143 & 1596 $\pm$ 1263 & 2.71 $\pm$ 0.76 & 44.909 $\pm$ 0.122 & 48.5\\
6 & Sh2-255 IR           & 500 $\pm$ 50 & 0.26 $\pm$ 0.08 & 500 $\pm$ 255 & 159 $\pm$ 83 & 833 $\pm$ 395 & 34.1 $\pm$ 3.5 & 46.048 $\pm$ 0.045 & 55.0 \\
7 & [IBP2002] CC14    & 47 $\pm$ 5 & 0.18 $\pm$ 0.08 & 63 $\pm$ 32 & 61 $\pm$ 40 & 419 $\pm$ 331 & 0.024 $\pm$ 0.002  & 42.822 $\pm$ 0.044 & 55.1\\
8 & IRAS 06058+2138   & 127 $\pm$ 16 & 0.18 $\pm$ 0.08 & 169 $\pm$ 87 & 140 $\pm$ 90 & 1476 $\pm$ 1161 & 1.10 $\pm$ 0.11 & 44.519 $\pm$ 0.044 & 96.1\\
9 & IRAS 06065+2124   & 43 $\pm$ 4 & 0.22 $\pm$ 0.10 & 29 $\pm$ 15 & 22 $\pm$ 15 & 160 $\pm$ 127 & 0.014 $\pm$ 0.001 & 42.679 $\pm$ 0.044 & 24.7\\
10 & NGC 2282         & 89 $\pm$ 13 & 0.34 $\pm$ 0.14 & 45 $\pm$ 23 & 20 $\pm$ 14 & 79 $\pm$ 70 & 10.30 $\pm$ 2.20 & 45.405 $\pm$ 0.093 & 2.1\\
11 & IRAS 05490+2658  & 220 $\pm$ 23 & 0.19 $\pm$ 0.09 & 440 $\pm$ 225 & 135 $\pm$ 87 & 508 $\pm$ 402 & 1.962 $\pm$ 0.838  & 44.889 $\pm$ 0.186 & 66.8\\
12 & BFS 56           & 70 $\pm$ 7  & 0.12 $\pm$ 0.06  & 70 $\pm$ 36 & 28 $\pm$ 18 & 361 $\pm$ 285 & 0.036 $\pm$ 0.004 & 43.156 $\pm$ 0.044 & 106.4\\
%13 & [BDS2003] 89  & 3.0 & 110 $\pm$ 14 & 0.62 $\pm$ 0.10 & 37 $\pm$ 5 & 69 $\pm$ 8.8 & 242 $\pm$ 103 & 3.84 $\pm$ 0.384 & 44.808 $\pm$ 0.044 & 2.37\\
13 & Sh2-88           & 122 $\pm$ 11 & 0.17 $\pm$ 0.06 & 407 $\pm$ 207 & 444 $\pm$ 228 & 2442 $\pm$ 1096 & 4189.0 $\pm$ 59.0 & 48.160 $\pm$ 0.006 & 3.6 \\
14 & IRAS 06103+1523  & 72 $\pm$ 8 & 0.14 $\pm$ 0.07 & 58 $\pm$ 30 & 87 $\pm$ 56 & 2750 $\pm$ 2176 & 1.0 $\pm$ 0.1 & 44.588 $\pm$ 0.044 & 99.2 \\ 
15 & IRAS 06104+1524A & 83 $\pm$ 9 & 0.33 $\pm$ 0.12 & 83 $\pm$ 43 & 125 $\pm$ 80 & 654 $\pm$ 518 & 5.73 $\pm$ 0.75 & 45.291 $\pm$ 0.057 & 7.7 \\
16 & IRAS 06117+1901  & 245 $\pm$ 26 & 0.28 $\pm$ 0.11 & 163 $\pm$ 84 & 20 $\pm$ 13 & 61 $\pm$ 48 & 14.34 $\pm$ 1.43 & 45.772 $\pm$ 0.044 & 5.8 \\
17 & IRAS 05480+2545  & 81 $\pm$ 9 & 0.07 $\pm$ 0.04 & 108 $\pm$ 55 & 54 $\pm$ 35 & 1543 $\pm$ 1221 & 0.54 $\pm$ 0.05 & 44.363 $\pm$ 0.044 & 236.1\\
\hline
\hline
\end{tabular}

%\end{table*}

\end{sideways}
\caption{Cluster properties.} 
\label{tab:clusters:out}
\end{table}

\subsection{Scaling relations at clump scale}

\begin{figure*}
    \centering
    \includegraphics[width= 15cm]{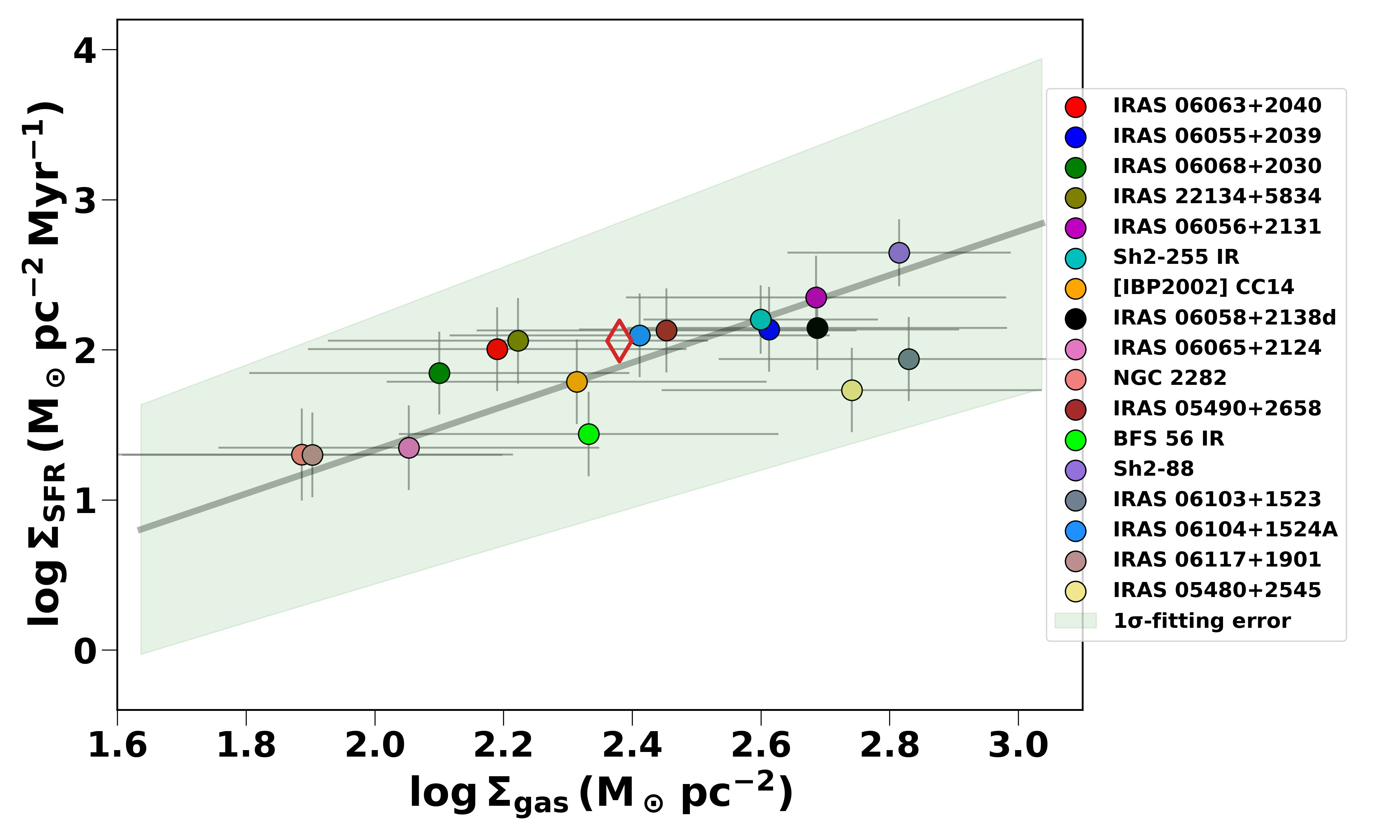}
    \caption{Variation of $\log \Sigma_{\rm{SFR}}$ with $\log \Sigma_{\rm{gas}}$. The black line shows the ODR fit with a best-fit power law exponent of $\sim$1.5 along with 1$\sigma$ uncertainty shown as a green shaded region. The diamond symbol denotes the FSR 655 cluster studied in \citet{Rawat_2024c}. }
    \label{fig:SFR-gas_mass}
\end{figure*}

Observational studies show that the number of young stars is correlated with the gas density, which means that most of the YSOs form in the higher-density structures of the cloud \citep{Heiderman_2010, Lada_2012, Lada_2013, Evans_2014}. %\textbf{However, once the feedback effect from massive stars becomes dominant in a star-forming region, it can promote the formation of new stars and also can disperse the natal gas material as it evolves. \textcolor{red}{this sentence looks odd here - not connecting well next sentences}} 
Therefore, to investigate the relation between the SFR and gas mass, we determined $\Sigma_{\mathrm{SFR}}$ and $\Sigma_{\rm{gas}}$ by dividing the SFR and gas mass from the projected area ($A_{\rm{clust}}$ = $\pi R^2_{\rm{clust}}$) of the clumps on the plane of the sky. The $\Sigma_{\rm{SFR}}$ and $\Sigma_{\rm{gas}}$ values for all the clusters %, along with their uncertainties, 
are given in Table \ref{tab:clusters:out}. The mean and median $\Sigma_{\rm{SFR}}$ of the clusters in our sample are $\sim$114 and $\sim$101 \Ms~Myr$^{-1}$ pc$^{-2}$, respectively, while the mean and median $\Sigma_{\rm{gas}}$ are $\sim$314 and $\sim$258 \Ms~pc$^{-2}$, respectively. Fig. \ref{fig:SFR-gas_mass} illustrates the variation of $\Sigma_{\rm{SFR}}$ with $\Sigma_{\rm{gas}}$, which shows a positive correlation. The Pearson correlation coefficient in the log-log scale is around 0.73. %Baring the outlier, IRAS 05480$+$2545 (yellow dot in Fig. \ref{fig:SFR-gas_mass}), we find that the Pearson correlation coefficient comes around 0.83. 
As discussed earlier, scaling relations follow a power-law relation ($\Sigma_{\rm{SFR}} = A \Sigma_{\rm{gas}}^N$), which in the logarithmic form can be expressed as

\begin{equation}
    \log \left(\frac{\Sigma_{\rm{SFR}}}{\rm{1 \Ms~Myr^{-1}pc^{-2}}}\right) = \log A + N \log \left(\frac{\Sigma_{\rm{gas}}}{\rm{1 \Ms~pc^{-2}}}\right).
\end{equation}\\

%We first did the least-squares fitting of the data with the above equation by considering the errors only in $\Sigma_{\rm{SFR}}$. From the weighted fit, we find the best-fit value of $N$ to be $\sim$xx $\pm$ xx and the normalization constant, $A$, to be $\sim$xx $\pm$ xx. 
\noindent To consider the errors in both axes, we adopted the Orthogonal Distance Regression (ODR) method \citep{Boggs_1988} to fit the data points. With ODR, the best-fit $N$ tuns out to
be $\sim$1.46 $\pm$ 0.28.
%With ODR, the best-fit $N$ and $\log A$ values were found to be $\sim$1.46 $\pm$ 0.28 and $-$ 1.58 $\pm$ 0.69, respectively. 
In Fig. \ref{fig:SFR-gas_mass}, we also show the position of the young cluster FSR 655, studied by \cite{Rawat_2024c} using a method similar to the one employed in this work.

\begin{figure*}
    \centering
    \includegraphics[width= 15cm]{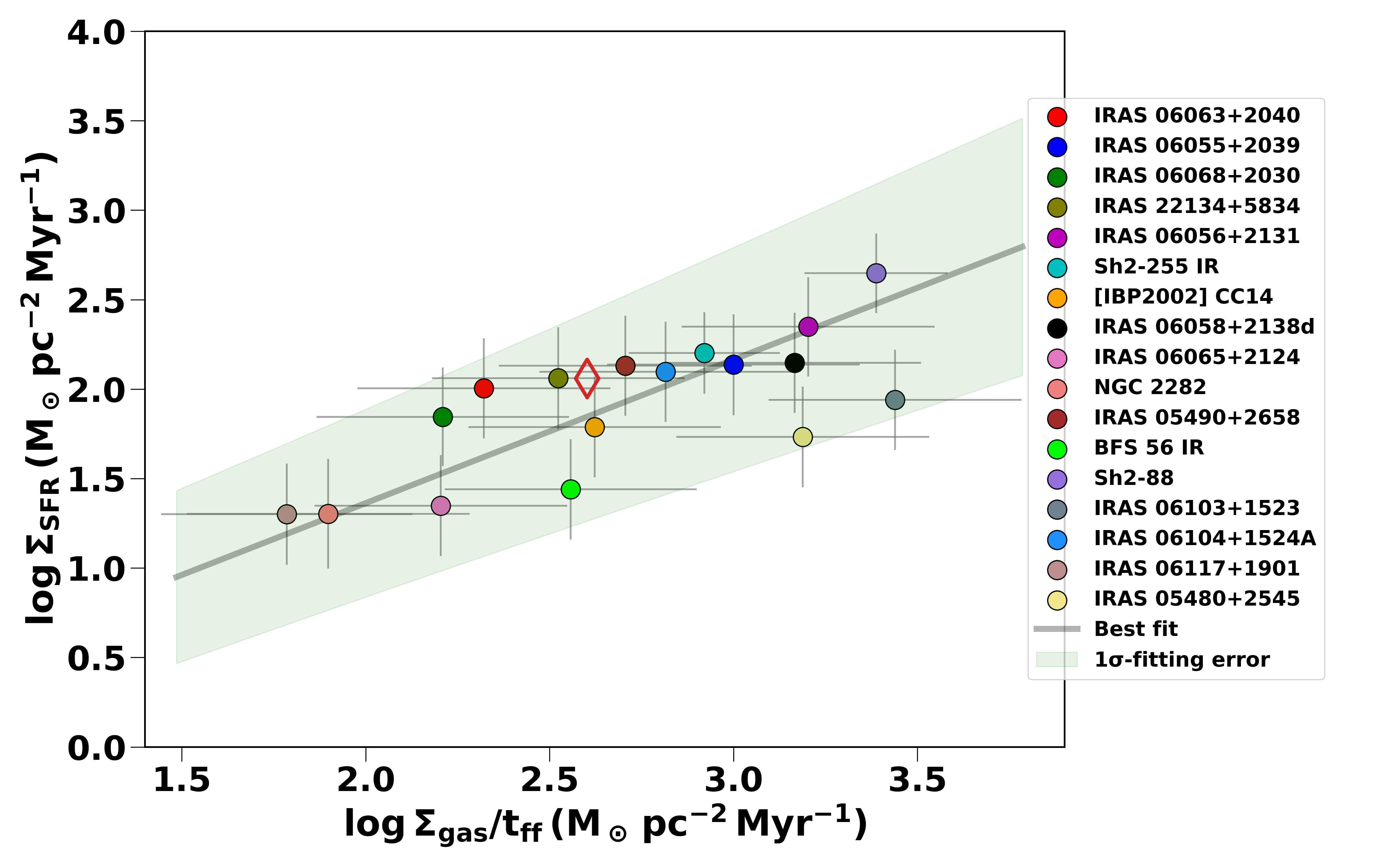}
    \caption{Variation of $\log \Sigma_{\rm{SFR}}$ with $\log \Sigma_{\rm{gas}}/t_{\rm{ff}}$. The black line shows the ODR fit with a best-fit power law exponent of $\sim$0.8 along with 1$\sigma$ uncertainty shown as a green shaded region. The diamond symbol is the same as in Fig. \ref{fig:SFR-gas_mass}.}
    \label{fig:SFR-gas_mass_tff}
\end{figure*}

The theoretical models \citep{Krum_mckee_2005, Krumholz_2019} predict the dependence of star formation rate on the free-fall time of the clouds. It has also been found observationally that the inclusion of the free-fall time reduces the scatter in the SFR$-$gas mass relation \citep{Krumholz_2012, Riwaj_2021}. This relation is the volumetric star formation relation, which in logarithmic form can be expressed as 

\begin{multline}
\label{vol_sfr_rel}
    \log \left(\frac{\Sigma_{\rm{SFR}}}{\rm{1 \Ms~Myr^{-1}pc^{-2}}}\right) = \log A^\prime\\ + N^\prime \log \left(\frac{\Sigma_{\rm{gas}}/t_{\rm{ff}}} {\rm{1 \Ms~Myr^{-1}pc^{-2}}}\right),
\end{multline}\\
\noindent where $t_{\rm{ff}}$ depends only on the volume density of the region. For the studied sample,
we found that 
%We also explored the volumetric star formation relation for the sample of clumps studied in this work. 
the $\Sigma_{\rm{gas}}/t_{\rm{ff}}$  lies in the range of $\sim$61 to 2750 \Ms~Myr$^{-1}$ pc$^{-2}$ with a mean and median around 858~\Ms~Myr$^{-1}$ pc$^{-2}$ and 508 \Ms~Myr$^{-1}$ pc$^{-2}$, respectively. Fig. \ref{fig:SFR-gas_mass_tff} shows the $\Sigma_{\rm{SFR}}$ vs $\Sigma_{\rm{gas}}/t_{\rm{ff}}$ plot, %which shows less scatter in comparison to $\Sigma_{\rm{SFR}}$ vs $\Sigma_{\rm{gas}}$ plot and a positive correlation 
which are well correlated with a Pearson's coefficient of $\sim$0.74. %Again, baring IRAS 05480$+$2545 cluster in the plot, we find that Pearson's correlation coefficient changes to $\sim$0.83. 
We fitted the relation with equation \ref{vol_sfr_rel} using the ODR method and found the best-fit value of $N^\prime$ 
as $\sim$0.80 $\pm$ 0.15. 
%and $\log A^\prime$ to be $\sim$0.80 $\pm$ 0.15 and $-$ 0.24 $\pm$ 0.43, respectively.
The exponent value obtained for clumps matches well, within the uncertainty, with the mean and median exponent values reported by \cite{Riwaj_2021} (i.e. $\sim$0.94 and $\sim$0.99, respectively) for volumetric star formation relation at the cloud scale.%, as they have also used the ODR method for fitting (\textcolor{red}{what is their error values in N and A??. May be you can also give}).\\ 

\section{Discussion}
\label{discuss_1}

\subsection{Comparison with existing star formation scaling relations}

As already discussed, several previous studies investigated the scaling relations at cloud scale or within single clouds \citep{Evans_2009, Lada_2010, Heiderman_2010, Gutermuth_2011, Krumholz_2012, Evans_2014, Hony_2015, Nalin_2016, Das_2021, Riwaj_2021}. It is important to acknowledge that the shapes of the scaling relations are sensitive to the adopted methodologies (see Table \ref{tab:scaling_summary}) due to systematic uncertainties and data quality. Therefore, we do not intend to directly compare the derived scaling relations with those from previous studies but instead want to analyze them in the context of the trends observed in the aforementioned studies. These studies have been done over the nearby clouds ($<$ 1 kpc) and are based on the star count method to determine the SFR. The aforementioned studies tested scaling relations in various forms and found different power-law exponents, which, up to some extent, can be attributed to the differences in methodology, like data resolution, SFR tracers, fitting methods, gas tracers, and completeness of the YSO sample. Moreover, some of the studies found that there is relatively large scatter and loose correlation in the SFR$-$gas mass relation between clouds in comparison to within single clouds \citep[e.g.][]{Lada_2013, Evans_2014}.

\begin{figure*}
    \centering
    \includegraphics[width= 15cm]{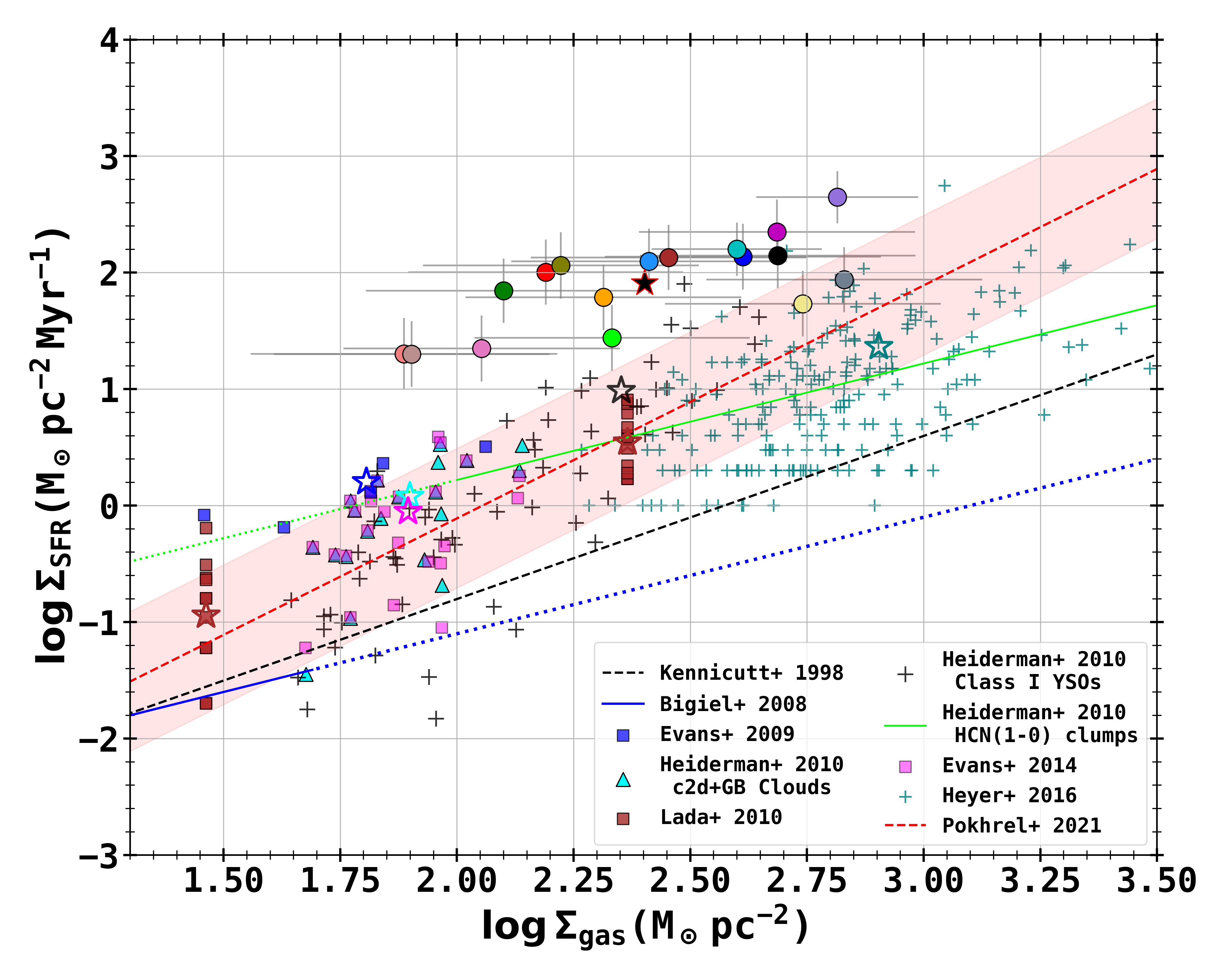}
    \caption{Comparison of the $\Sigma_{\rm{SFR}}-\Sigma_{\rm{gas}}$ relation obtained in this work with existing relations in the literature. The solid colored dots denote the clusters in our sample, as shown in Fig. \ref{fig:SFR-gas_mass}. The galactic-scale relations of \citet{Kennicutt_1998b} and \citet{Bigiel} are shown by black dashed and blue solid lines, respectively. The \citet{Bigiel}'s relation is extrapolated by a blue dotted line towards higher gas mass surface densities. The nearby clouds from \citet{Evans_2009} (blue squares), \citet{Heiderman_2010} (cyan triangles), \citet{Lada_2010} (brown squares), and \citet{Evans_2014} (pink squares) are shown. The black pluses show the $\Sigma_{\rm{SFR}}-\Sigma_{\rm{gas}}$ values obtained for only Class I YSOs in the nearby clouds by \citet{Heiderman_2010}. The green solid line shows the relation obtained for HCN (1--0) massive dense clumps by \citet{Heiderman_2010}, which is extrapolated by a green dotted line towards lower gas mass surface densities. The teal pluses show the massive clumps from \citet{Heyer_2016}. The red dashed line shows the $\Sigma_{\rm{SFR}}-\Sigma_{\rm{gas}}$ relation obtained by \citet{Riwaj_2021}, with a spread shown as a red shaded area (see text for details). The mean $\Sigma_{\rm{SFR}}-\Sigma_{\rm{gas}}$ in our sample is shown by a black solid star, and for other samples, the mean values are shown by open stars of the same colors as their corresponding sample.}

    \label{fig:SFR-Mg_comp}
\end{figure*}

In Fig. \ref{fig:SFR-Mg_comp}, a comparison of the results of this work with the existing scaling relations at the extragalactic and cloud/clump scale are shown. From the figure, it can be seen that although the value of the exponent obtained in this work ($\sim$1.5) is similar to the KS power law index \citep [$\sim$1.4;][]{Kennicutt_1998b}, the data points from our sample of clumps lie well above the KS relation. Fig. \ref{fig:SFR-Mg_comp} also shows the linear relation between $\Sigma_{\rm{SFR}}$ and molecular gas surface density obtained by \cite{Bigiel} for 18 nearby galaxies at $\sim$750 pc scales, with $\Sigma_{\rm{H_2}}$ in the range of 3$-$50 \Ms~pc$^{-2}$. %, which is shown by a solid blue line and extrapolated by a blue dotted line towards the higher gas densities. 
The data points of this work also show much higher $\Sigma_{\rm{SFR}}$ values than predicted by the linear relation of \cite{Bigiel}. The higher trend of scaling relations than the extragalactic ones has also been found at the cloud scale by other observational studies \citep{Evans_2009, Lada_2010, Heiderman_2010, Hony_2015}. From the figure, one can also see that our studied sample shows higher
values in comparison to nearby clouds from c2d and GB survey \citep{Evans_2009, Lada_2010, Heiderman_2010, Evans_2014}.
%our cluster forming clumps lies above in the $\Sigma_{\rm{SFR}}-\Sigma_{\rm{gas}}$ plot (see Fig. \ref{fig:SFR-Mg_comp}). %However, the power-law index obtained for our sample of 17 cluster-forming clumps is shallower than the index value of \citep[$\sim$2.04;][]{Evans_2014}, but it is still consistent within the uncertainty. 
\cite{Heiderman_2010} also examined the scaling relation for the youngest YSOs (e.g. Class I) and found that their $\Sigma_{\rm{SFR}}$ is higher than the values obtained by including all the YSOs in the clouds, which is also shown in Fig. \ref{fig:SFR-Mg_comp}. From the figure, it can also be seen that though the Class I YSO sample of \cite{Heiderman_2010} is closest to our sample, still the clumps within our sample exhibit higher values of $\Sigma_{\rm{SFR}}$ compared to their Class I YSO sample. As discussed earlier, the GB survey traces low-mass star-forming clouds, which may influence the derived values of the SFRs and SFEs. \cite{Riwaj_2021} studied 12 nearby MCs and obtained the $\Sigma_{\rm{SFR}}-\Sigma_{\rm{gas}}$ relation within individual clouds, with a mean and median power-law exponent of $\sim$2.00 and $\sim$2.08, respectively, and a cloud-to-cloud spread of $\sim$0.3 dex in $\Sigma_{\rm{SFR}}$ at logarithmic scale. The obtained power law exponent in this work is shallower, but still consistent within 2$\sigma$ uncertainty. %Also, \cite{Riwaj_2021} found a shallower index value of $\sim$1.44 $\pm$ 0.06 for AFGL 490. 
However, as can be seen from Fig. \ref{fig:SFR-Mg_comp}, our cluster sample lies above the \cite{Riwaj_2021} scaling relation. As discussed in \citet{Suin_2024}, the sensitivity and resolution of the data influence the determination of the star formation properties and, consequently, the scaling relations.  This is particularly true in the cluster environment, where the bright background affects the detection of the faint point sources 
in the $\it{Spitzer}$ bands compared to NIR bands \cite[e.g. see Figure 13 of][]{Rawat_2024c}.
%Taking a mean $\Sigma_{\rm{gas}}$ of the clumps to be $\sim$282 \Ms~pc$^{-2}$, the predicted value of $\Sigma_{\rm{SFR}}$ from the scaling relation of \cite{Riwaj_2020, Riwaj_2021} comes around 6.2 \Ms~pc$^{-2}$ Myr$^{-1}$, which is around 20 times lower than the corresponding value from the obtained scaling relation in this work. 
%The primary explanation for the higher $\Sigma_{\rm{SFR}}$ values in our case is likely the chosen scale size of our sample, $\sim$1$-$2 pc, and also we have accounted for all stars down to 0.1 \Ms~while estimating cluster properties. 
%The better sensitivity and resolution of UKIDSS data can trace young stars deeper than Spitzer data, which is used in \cite{Riwaj_2021}. As discussed, these factors influence SFRs and, consequently, the scaling relations \citep{Suin_2024}.}  %The $\Sigma_{\rm{SFR}}$ and $\Sigma_{\rm{gas}}$ of some Galactic high-mass star-forming molecular clouds studied by \cite{Romero_2017} are also shown in the figure, which have high $\Sigma_{\rm{SFR}}$ values but still lower than our sample of clumps.    

At the clump scale, \cite{Heiderman_2010} investigated the SFR$-$gas mass relation in massive dense clumps (mean radius $\sim$1.13 pc) from \cite{Wu_2010} that are traced by HCN (1$-$0) molecular line data. The authors calculated the gas mass from HCN and SFR from infrared luminosities (8$-$1000 \mum) and obtained a linear dependence of $\Sigma_{\rm{SFR}}$ on $\Sigma_{\rm{HCN}}$, which is also shown %by a cyan solid line and extrapolated towards higher gas densities by a dashed cyan line 
in Fig. \ref{fig:SFR-Mg_comp}. From the figure, it can be seen that our cluster sample lies above their relation. %by a factor of $\sim$25. 
In this case, more than half of the sample of \cite{Wu_2010} consists of clumps beyond 4 kpc. %and the sample has been studied using infrared luminosities. 
At large distances, some faint low-mass stars might be missed, leading to an underestimation of the SFR. %Also, our clumps show much higher values of $\Sigma_{\rm{SFR}}$ and less scatteredness than the clumps of NAN-complex (North American Nebula and the Pelican Nebula) studied by \cite{Das_2021}. 
The massive clumps from the study of \cite{Heyer_2016} are also shown in the figure, and it can be seen that few of them lie close to our cluster sample. \cite{Heyer_2016} studied star formation scaling relations in massive clumps of the Milky Way by selecting the clumps from APEX Telescope Large Area Survey of the Galaxy \citep[ATLASGAL;][]{Schuller_2009} data and linking them to the YSOs from the catalogue of $\it{Spitzer}$ 24 \mum MIPSGAL survey. The authors calculated the total gas mass from the 870 \mum flux and evaluated the total stellar mass by sampling the IMF and computed the SFR by dividing the inferred stellar mass by time scales of 0.5 Myr.  Our sample shows higher
values in comparison to clumps studied by \cite{Heyer_2016}. %Although the exact reason is not known to us, however, it is worth noting that due to the low sensitivity of the MIPSGAL 24 \mum data, the mass sensitivity limit of the \cite{Heyer_2016}'s YSO sample is only down to 2 \Ms. 
A possible explanation is 
that due to the lower sensitivity of the MIPSGAL 24 \mum data, the mass sensitivity limit of the \cite{Heyer_2016}'s YSO sample is only down to 2 \Ms. In most of the clumps, they detected only one or a few protostars in 24 \mum, which might have added uncertainty in estimating the total stellar mass of the clumps in their study. In addition, \cite{Heyer_2016}'s 
clumps are likely at earlier evolutionary stages compared to
the clumps studied in this work. 
We want to emphasize
that, although the NIR data used in this work are largely complete down to 0.2\Ms, however,
%The primary explanation for the higher $\Sigma_{\rm{SFR}}$ values in our case is likely the chosen scale size of our sample, $\sim$1$-$2 pc, and also 
we have accounted for all the stars down to 0.1 \Ms~by extrapolating the IMF while estimating the clusters' properties. 

%for most of the clusters, data is sensitive down to 0.2 \Ms~limit. %However, it is to be noted that the sample in \cite{Heyer_2016} represents very early stages of cluster-forming clumps, whereas the cluster sample studied in this work is relatively more evolved, with a cluster visibly emerging from the clump in NIR.

%The $\Sigma_{\rm{SFR}}-\Sigma_{\rm{gas}}$ relation for our cluster-forming clumps shows better correlation and less scatteredness in comparison to those found in some of the earlier studies at cloud/clump scale \citep{Heiderman_2010, Lada_2013, Evans_2014, Nalin_2016, Heyer_2016, Romero_2017}. 
%{\bf This scatteredness in our sample of clusters can be attributed to the narrow evolutionary stages of our sample as well as to their properties being derived over
%smaller spatial scales.} The clusters which are at the initial stages of star formation are gas-rich, while the relatively older clusters tend to deplete gas due to ongoing star formation and associated feedback, as also highlighted by \cite{Megeath_2022} \textcolor{red}{This para not needed as we are discussing
%this again in the next to next para}.  

\begin{figure*}
    \centering
    \includegraphics[width= 15cm]{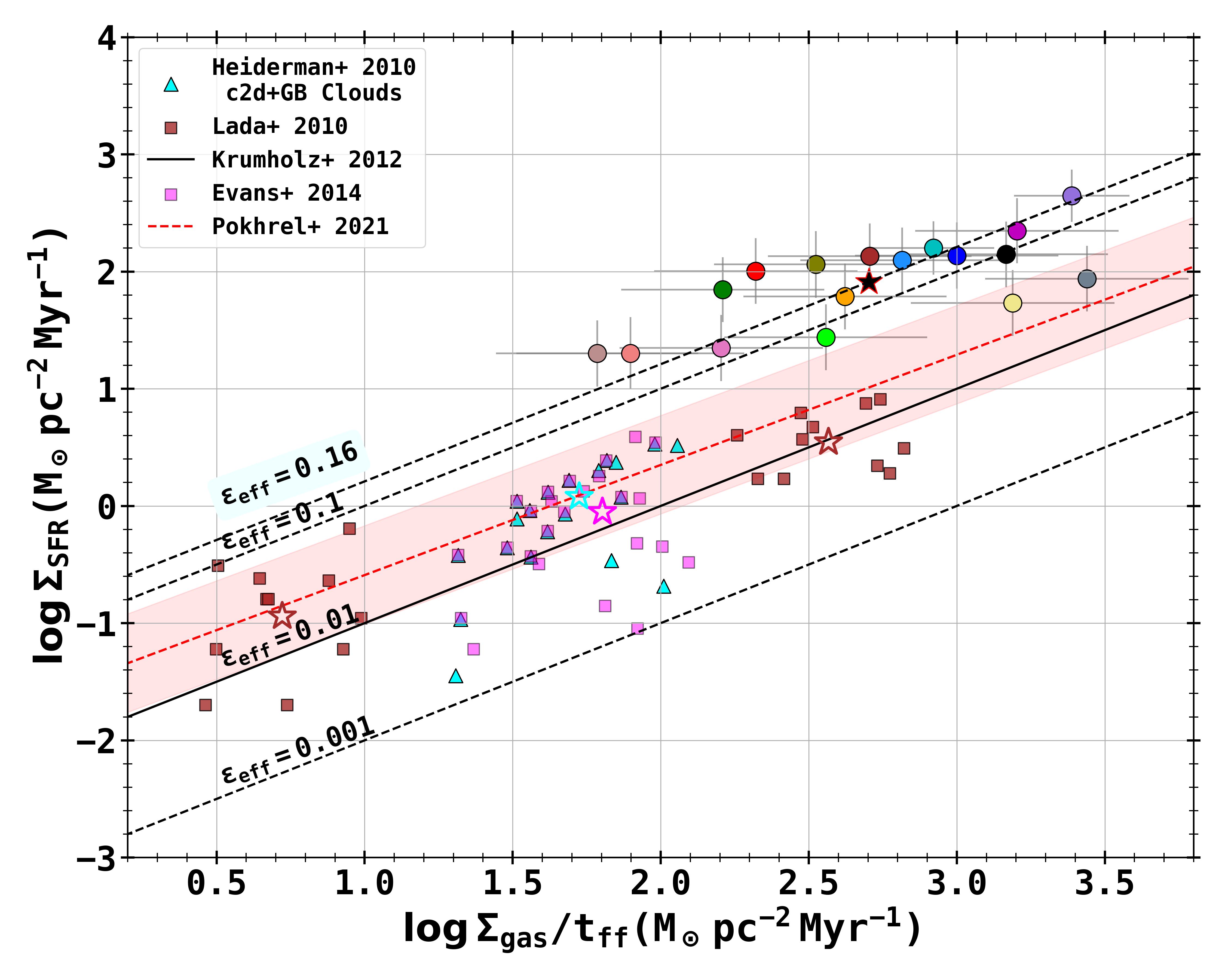}
    \caption{Comparison of $\log \Sigma_{\rm{SFR}}$ with $\log \Sigma_{\rm{gas}}/t_{\rm{ff}}$. The coloured symbols, red dashed line and red shaded region are the same as in Fig. \ref{fig:SFR-Mg_comp}. The black solid line shows the relation of \citet{Krumholz_2012} and denotes the $\epsilon_{\rm{ff}}$ of 0.01, and the black dashed lines show the $\epsilon_{\rm{ff}}$ of 0.001, 0.1, and 0.16, respectively.}
    \label{fig:SFR-Mg_comp2}
\end{figure*}

Fig. \ref{fig:SFR-Mg_comp2} shows the comparison of the volumetric star formation relation obtained for cluster sample in this work with those observed in some of the previous studies at the cloud scale. Similar to the $\Sigma_{\rm{SFR}}-\Sigma_{\rm{gas}}$ plot, it is clearly evident from the $\Sigma_{\rm{SFR}}-\Sigma_{\rm{gas}}/t_{\rm{ff}}$ plot that our cluster sample lies above the nearby clouds of previous studies \citep{Heiderman_2010, Lada_2010, Evans_2014, Riwaj_2021}. %Also, our work shows less scatteredness than all aforementioned studies except \cite{Riwaj_2021}. 
The obtained slope of the volumetric star formation relation for the cluster sample of this work is almost the same as found in \cite{Riwaj_2021} (i.e. 0.94); however, it is apparent from the figure that the trend line for our cluster sample lies above that of \cite{Riwaj_2021}'s relation. \cite{Krum_mckee_2005} theoretically derived the value of $\epsilon_{\rm{ff}}$ to be $\approx$ 0.01 for any supersonically turbulent medium. At large scales (length $\sim$100 pc) also, the $\epsilon_{\rm{ff}}$ is found to be $\approx$ 0.01 in nearby galaxies using CO molecular data and IR luminosities \citep[see][and references therein]{Chevance_2023}. For Galactic MCs, nearby galaxies, and high redshift galaxies, \cite{Krumholz_2012} also observed a linear relation between $\Sigma_{\rm{SFR}}$ and $\Sigma_{\rm{gas}}/t_{\rm{ff}}$ along with a $\epsilon_{\rm{ff}}$ value of $\sim$0.01, which they suggested would be roughly constant with a dispersion of $\sim$0.3 dex \citep[see also Figure 10 of][]{Krumholz_2019}. The study by \cite{Riwaj_2021} in nearby clouds, found $\epsilon_{\rm{ff}}$ in the range of 0.010 to 0.043, with a median around 0.026, and did not find any threshold density above which the $\epsilon_{\rm{ff}}$ rises significantly. Fig. \ref{fig:SFR-Mg_comp2} shows that although the best-fit slope ($\sim$0.8) found here fairly matches the linear relation of \cite{Krumholz_2012} (shown by a solid black line) within the uncertainty, the $\epsilon_{\rm{ff}}$ for our cluster sample is higher than their obtained value of 0.01. While the mean $\epsilon_{\rm{ff}}$ value for our cluster sample is $\sim$0.15, fixing the slope of $\Sigma_{\rm{SFR}}-\Sigma_{\rm{gas}}/t_{\rm{ff}}$ relation to unity (i.e. $\Sigma_{\rm{SFR}} \propto {\Sigma_{\rm{gas}}/t_{\rm{ff}}}^{1.0}$), the best-fit $\epsilon_{\rm{ff}}$ value from the ODR regression fit comes around 0.16 (shown by a dashed black line in Fig. \ref{fig:SFR-Mg_comp2}), which is 16 times higher than the theoretical $\epsilon_{\rm{\rm{ff}}}$ value of $\sim$0.01 \citep{Krum_mckee_2005}. %\cite{Riwaj_2021} in their study of nearby clouds also tested the volumetric star formation relation for individual clouds, and found best-fit slopes close to unity for each cloud, with a mean around 0.94 and a spread of 0.21 in log $\Sigma_{\rm{SFR}}$. The relation of \cite{Riwaj_2021} is also shown in Fig. \ref{fig:SFR-Mg_comp2} by a dashed red line. 

%\textcolor{red}{Overall, it is apparent that our cluster sample shows significantly higher star formation rate surface densities than most of those found by previous studies (see Fig. \ref{fig:SFR-Mg_comp} and \ref{fig:SFR-Mg_comp2})}. 
Although the power law exponent in $\Sigma_{\rm{SFR}}-\Sigma_{\rm{gas}}$ relation for our cluster sample is somewhat comparable with the previous studies within the uncertainty, the surface density values of the SFR at the same gas mass surface densities are relatively higher. These high values of $\Sigma_{\rm{SFR}}$ for our cluster sample can be explained by the following reasons: (i) the regions studied in this work are relatively much smaller than those at extragalactic scales \citep{Kennicutt_1998b, Bigiel}, (ii) our sample consists of cluster forming regions, i.e. active star-forming regions; however, at cloud and mainly extragalactic scales, averaging over large regions includes both active and quiescent (i.e. diffused non-star forming gas) star-forming regions. As a consequence, regions in our sample have higher SFRs within smaller plane-of-sky projected surface areas, hence larger $\Sigma_{\rm{SFR}}$ values. A less scatter in the SFR$-$gas mass relation found here (see Fig. \ref{fig:SFR-Mg_comp} and \ref{fig:SFR-Mg_comp2}) can be explained by the adopted methodology for calculating the total stellar mass and the narrow evolutionary spread of the
clumps, i.e. the ages of most of the embedded clusters are in the range 0.5$-$1.5 Myr. In this work, to get the total stellar mass, we extrapolated the IMF down to the low mass limit of $\sim$0.1 \Ms~and calculated the ages of the clusters as a whole by comparing them with the synthetic clusters of different ages. While, in some of the previous studies, a single average mass ($\sim$0.5 \Ms) and age ($\sim$2 Myr) for all YSOs were adopted \citep{Evans_2009, Lada_2010, Heiderman_2010, Evans_2014}. To avoid the uncertainty caused by including the Class II YSOs, some of the previous studies included only young protostars (e.g. Class I) and found relatively better correlation between $\Sigma_{\rm{SFR}}$ and $\Sigma_{\rm{gas}}$ \citep{Heiderman_2010, Lombardi_2013, Lada_2013, Evans_2014, Willis_2015}. 
%While in this work, we have estimated the total stellar mass of the clusters by extrapolating IMF down to 0.1 \Ms~and obtained their ages based on KLF modelling. 

%included all the field-subtracted cluster members to calculate the cluster properties, like SFR and SFE. %\textcolor{blue}{It is to be noted that most of the earlier studies based on the star-count method used $\it{Spitzer}$ data; however, recent JWST studies revealed a higher population of YSOs which were not found in earlier $\it{Spitzer}$ data. This higher number of YSOs will impact the overall SFE and SFRs. Therefore, the SFEs of the star-forming regions need to be reexamined using high sensitive UKIRT and JWST data \cite[e.g.][]{jones23}.} 

Apart from the above facts, in our cluster sample, except for NGC~2282 and IRAS 06117$+$1901, all other clusters have gas mass surface densities greater than 110 \Ms~pc$^{-2}$. A density threshold of $\sim$110$-$130 \Ms~pc$^{-2}$ (or $\rm{7-8A_V}$) has been suggested in the literature above which the SFR varies
linearly with the mass of dense gas and is better correlated \citep{Lada_2010, Heiderman_2010, Evans_2014, Das_2021, Patra_2022}. In fact, 14 out of 17 clusters in our sample have gas mass surface densities $\gtrsim$ 130 \Ms~pc$^{-2}$, which could also be a reason for a better correlation of the SFR with gas mass in the present work. Simulations also suggest a better correlation between SFR and gas mass at higher threshold density as a consequence of larger gravitational influence \citep{Burk_2013, soko19}. %\cite{Evans_2014} also mentioned that the threshold density, as indicated by \cite{Heiderman_2010} and \cite{Lada_2012}, is just a limit above which SFR becomes linear and better correlated to gas mass density, it is not a limit for star formation to occur. The authors also suggested that a particular threshold applicable to nearby clouds may not necessarily apply in other regions, like in extreme conditions of the central molecular zone \citep[for details, see][]{Longmore_2013} or low metallicity regions.    

\begin{figure}
    \centering
    \includegraphics[width= 8.5cm]{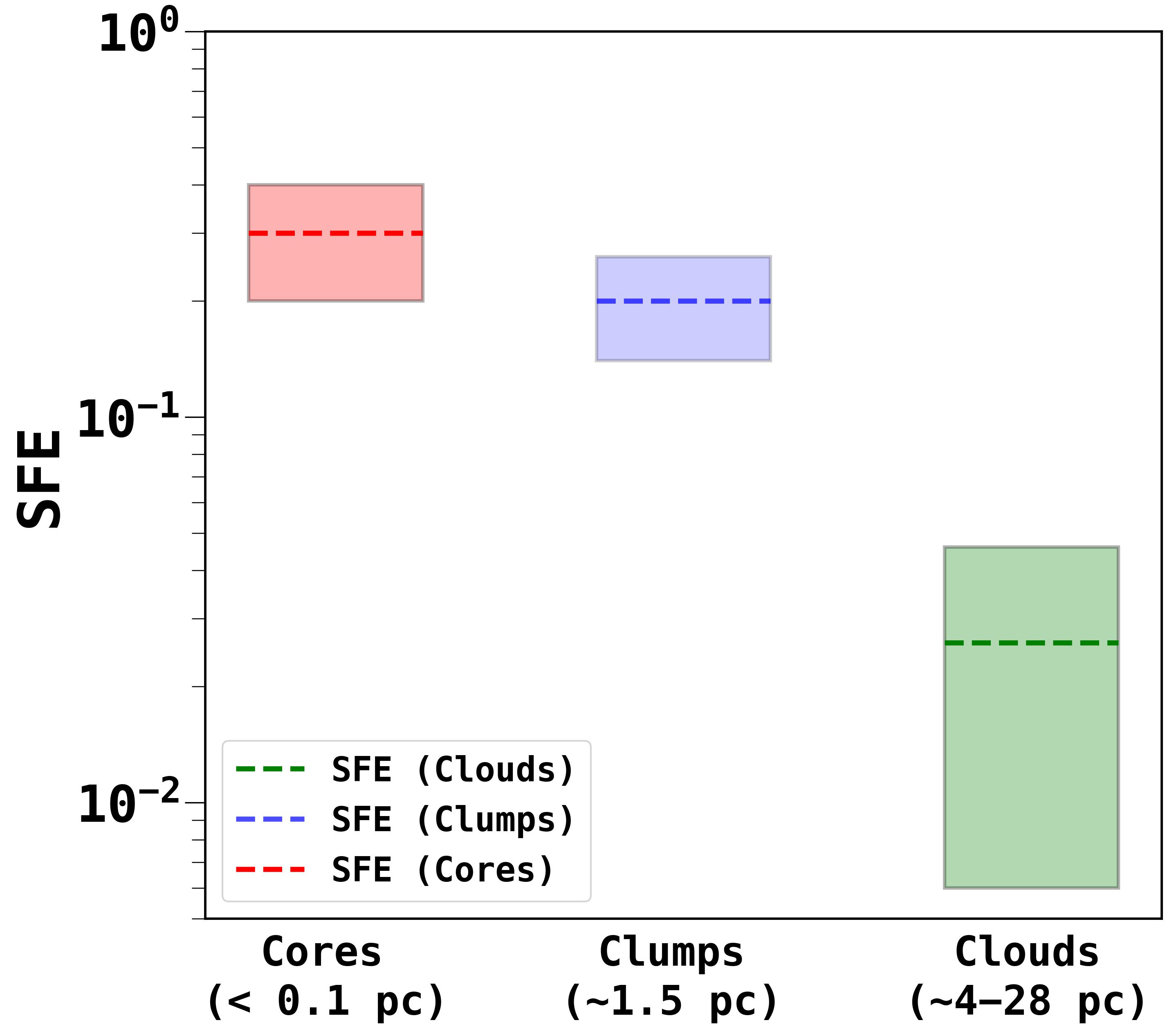}
    \caption{Star formation efficiencies at cloud, clump, and core scales. The red line shows the SFE of 30 $\pm$ 10\% at the core scale \citep{Alves_2007, Kon_2010, Kon_2015, Francesco_2020, Zhou_2024a}, the blue line shows the SFE of 20 $\pm$ 6\% at the clump scale found in this work, and the green line shows the SFE of 2.6 $\pm$ 2.0\% at the cloud scale \citep{Evans_2009, Heiderman_2010, Lada_2010, Evans_2014}. The corresponding colored shaded regions show the upper and lower limits of SFE at the respective scales. The mentioned range of cloud sizes is taken as the mean values from Table \ref{tab:scaling_summary}. }
    \label{fig:sfe_comp}
\end{figure}

Fig. \ref{fig:sfe_comp} shows the plot of average SFEs with the gas mass surface densities of clouds, clumps, and cores. The mean SFE at the cloud scale is taken from the studies of nearby clouds \citep{Evans_2009, Heiderman_2010, Lada_2010, Evans_2014}, which are based on the star count method. %We have also included the mean SFEs of Milky Way giant molecular clouds from \cite{Lee_2016} and high mass star-forming molecular clouds from \cite{Romero_2017}, which have traced the SFR from $\it{WMAP}$ free-free fluxes and mass-luminosity relation or sampling the IMF, respectively. 
From these studies, the SFE at the molecular cloud scale turns out to be around 2.6 $\pm$ 2.0\%. At the clump scale, the median SFE is around  20 $\pm$ 6.0\%, as obtained in this work from a sample of 17 cluster-forming clumps. 
%While the median standard deviation is around 8\%. 
At the core scale, the SFEs are indirectly derived from the similarity between the shape of the dense core mass function and the stellar IMF  for a sample of 
nearby molecular clouds and massive star-forming regions \citep{Alves_2007, Kon_2010, Kon_2015, Francesco_2020, Zhou_2024a}. 
These studies find that there is a one-to-one correlation between core and stellar masses, and the core-to-star formation efficiency is around 30 $\pm$ 10\%. Understanding how SFE
evolves from cloud scale to core scale is of particular interest for the overall evolution of the molecular cloud. Our analysis reveals an increasing efficiency from cloud to core scales, consistent with the hierarchical cloud structure model, where efficiency increases with the average gas density from large-scale clouds down to dense cores \citep{Elmegreen_2008}. However, establishing a definitive relationship between different scales and SFE would require a diverse sample spanning a broad range of surface densities, spatial scales, and evolutionary stages of clouds.

\subsection{Possible reasons for high $\epsilon_{\rm{ff}}$ at clump scale}
\label{high_eff}
In a recent work, \cite{Zamora_2024} simulated the collapse of a hub-filament system cloud under self-gravity to test the star formation-gas mass relations and found a similar exponent value for $\Sigma_{\rm{SFR}}$--$\Sigma_{\rm{gas}}/t_{\rm{ff}}$ relation (i.e. $\sim$0.94), as found in this work. %The gravo-turbulent model assumes that the clouds are supported against collapse by supersonic turbulence, and thus have long lifetimes and low SFE \citep{Krum_mckee_2005, Balles_2007, krum-mck_2020, Riwaj_2021}. 
\cite{Zamora_2024} suggest that the long lifetimes of clouds and their low SFE may not necessarily result from turbulent support against the cloud collapse, as proposed in gravo-turbulent model \citep{Krum_mckee_2005, krum-mck_2020}, but could instead simply be a consequence of cloud collapse under gravity. The GHC \citep[][]{sema2019} model suggests that the cloud appears to have a long lifetime because the accretion flows from low- to high-density regions continuously replenish the gas material in high-density regions to form stars. In this scenario, the SFR does not remain constant but increases over time till the stellar feedback becomes dominant to disrupt the cloud material, and then the SFR decreases \citep{sema2019}. \cite{Zamora_2024} also suggest that the observationally obtained SFR$-$gas mass relations at cloud scales \citep{Riwaj_2021}, characterized by low and nearly constant values of $\epsilon_{\rm{ff}}$, can be explained within the framework of the GHC model. The reasons behind low $\epsilon_{\rm{ff}}$ could be the accelerating SFR in clouds and the small free-fall time of high-density regions where stars are forming in comparison to the typical formation time of stars (for e.g. $t$ = 2 Myr for up to Class II YSO stage) \citep[for more details, see][]{Zamora_2024}. Due to these factors, the ratios on which the $\epsilon_{\rm{ff}}$ depends, $\frac{t_{\rm{ff}}}{t}$ and $\frac{M_\star}{M_{\rm{gas}}}$, would become very small and hence the low $\epsilon_{\rm{ff}}$ values. The authors argue that due to the exponentially increasing SFR along with $r^{-2}$ density profile, the total stellar mass formed from dense gas always remains smaller than the total gas mass of the whole region over which $\epsilon_{\rm{ff}}$ is calculated.

However, we note that in our study, cluster ages, $t_{\rm{clust}}$, are determined as a whole using the KLFs. As a result, in most cases, $t_{\rm{clust}}$ is less than three times the $t_{\rm{ff}}$, unlike the simulation results of \citet{Zamora_2024}, where this ratio ranges from 12 to 17. Secondly, while it is true at the cloud scale that high-density star-forming gas comprises only a small fraction of the total cloud mass, the surface density of most clumps in our sample exceeds 110 \Ms~pc$^{-2}$. %is defined differently: we selected clump radii based on stellar density, calculating all parameters within this region. As discussed in Section \ref{discuss_1}, However, the surface density of most clumps in our sample exceeds 110 \Ms~pc$^{-2}$, 
Therefore, in this work, clump gas masses predominantly consist of dense gas, and as a result, the $\frac{M_\star}{M_{\rm{gas}}}$ ratio is not as low in our sample of clumps.  %\textcolor{red}{We acknowledge that taking the size of the cluster and clump to be the same may introduce a caveat regarding the overestimation of the cluster's gas surface density. Nonetheless, the above assumption and not taking a single age for all the clusters could be a possible reason for higher values of $\epsilon_{\rm{ff}}$ in this work. }

%\textcolor{blue}{As previously discussed, stellar feedback, particularly from massive stars, plays a crucial role in the star formation process \citep{Geen_2015, Krumholz_2019}. 

Like \cite{Zamora_2024}, other studies also suggest that the photoionization feedback and stellar winds from the massive stars, in general, disperse the cloud material; thereby,  significantly reduces the global star formation properties of a cloud \citep{Dib_2011, Fed_kle_2012, Dib_2013, dale14, Geen_2015, Krumholz_2019}, although studies also suggest that feedback can also enhance star formation in some local regions of molecular clouds \citep{zav10,cha11,deh12,sam18}. In a recent simulation, \cite{Suin_2024} show that $\epsilon_{\rm{ff}}$ varies with the gas mass surface density and the evolution of the star-forming region and is also influenced by the local environment, such as the presence of stellar feedback. 
%{\bf There are also studies that suggest stellar feedback may increase star formation in some local regions of large-scale molecular clouds by sweeping and compressing the surrounding gas. However, these are mostly found around classical \hii regions/bubbles of size a few parsecs, where the \hii region has already elapsed a significant amount of time to trigger a new generation of stars. In the present case,
%we are dealing with clumps that are still associated with dense gas, and the embedded clusters are young ($<$ 1.5 Myr). Thus, the feedback from massive stars and the resulting presence of second-generation stars in the vicinity of the studied clusters are likely minimal. If present, they would be at very early evolutionary stages to be
%detected in NIR. In fact, we did not find cold dust shells or bright-rimmed clouds, the signatures of feedback-driven structures, around our studied cluster, strengthening our hypothesis of lack of second-generation stars in the studied
%area.}
%The photoevaporation and ionizing radiation from massive stars can disperse the parent cloud or clump material, forming $\hii$ regions and thereby reducing the SFR over time \citep{Fed_kle_2012, sema2019}.
%Although most of the clusters in our sample are young, the young massive stars can still have a significant effect on the collapse of clumps. 
%a few massive YSOs can be present. 

The presence of massive stars and their effect can be traced by measuring the thermal free-free radio continuum emission in a given region. Therefore,  we obtained radio continuum images from the NRAO VLA Sky
Survey (NVSS; $\nu$ $\sim$1.4 GHz, beam $\sim$45\arcsec $\times$ 45\arcsec) and NRAO VLA Archive Survey (NVAS; $\nu$ $\sim$1.4 GHz, beam $\sim$45\arcsec $\times$ 45\arcsec) \citep{Condon_1998} to search for radio emission in our cluster sample (e.g. shown in  Fig. \ref{fig:iras_2040_radio} for IRAS 06063+2040).
%We prioritized using the the D configuration data at 1.4/1.46 GHz with a beam size of $\sim$45\arcsec $\times$ 45\arcsec. 
For clumps that have negligible emission at 1.4 GHz, we search for images at $\sim$5 GHz in the VLA archive\footnote{\href { http://www.vla.nrao.edu/astro/nvas/}{ http://www.vla.nrao.edu/astro/nvas/}}, observed with the VLA D-configuration wherever applicable, considering the extended nature of the sources. For sources in which fluxes at both 1.4 GHz and 5 GHz are available, we prefer to use 5 GHz fluxes to minimize the effect of optical depth \citep[e.g.][]{yad22} on the emission at these frequencies. 
%to get the integrated flux.} 
We then calculated the integrated flux ($S_{\rm{\nu}}$) of the radio-emitting regions using the Gaussian fitting tool in CASA. In our sample, we note that no radio emission was found for 5 sources at both 1.4 GHz and 5 GHz. In such cases, we
%took either the integrated flux from the literature or
estimated the integrated flux from the images within the cluster radius and considered these fluxes
as the upper limit. The non-detection of radio emissions from a few sources implies that they may be at an early evolutionary stage
of massive star formation. 
%of the 
%the uniform emission within the cluster region above 5$\sigma$ from the background. Here, $\sigma$ is the rms value of the background emission at 1.4 GHz. 
From the integrated flux, we estimated the Lyman continuum photons ($N_{\rm{Lyc}}$) associated with individual clusters, whose details are given in Appendix \ref{Ionized}. The $S_{\rm{\nu}}$ and $N_{\rm{Lyc}}$ values for the clusters are given in Table \ref{tab:clusters:out}.  The $\log N_{\rm{Lyc}}$ range of 42.7 to 48.2 suggests that the strength of feedback is different across our cluster sample.

\begin{figure}
    \centering
    \includegraphics[width= 8.5cm]{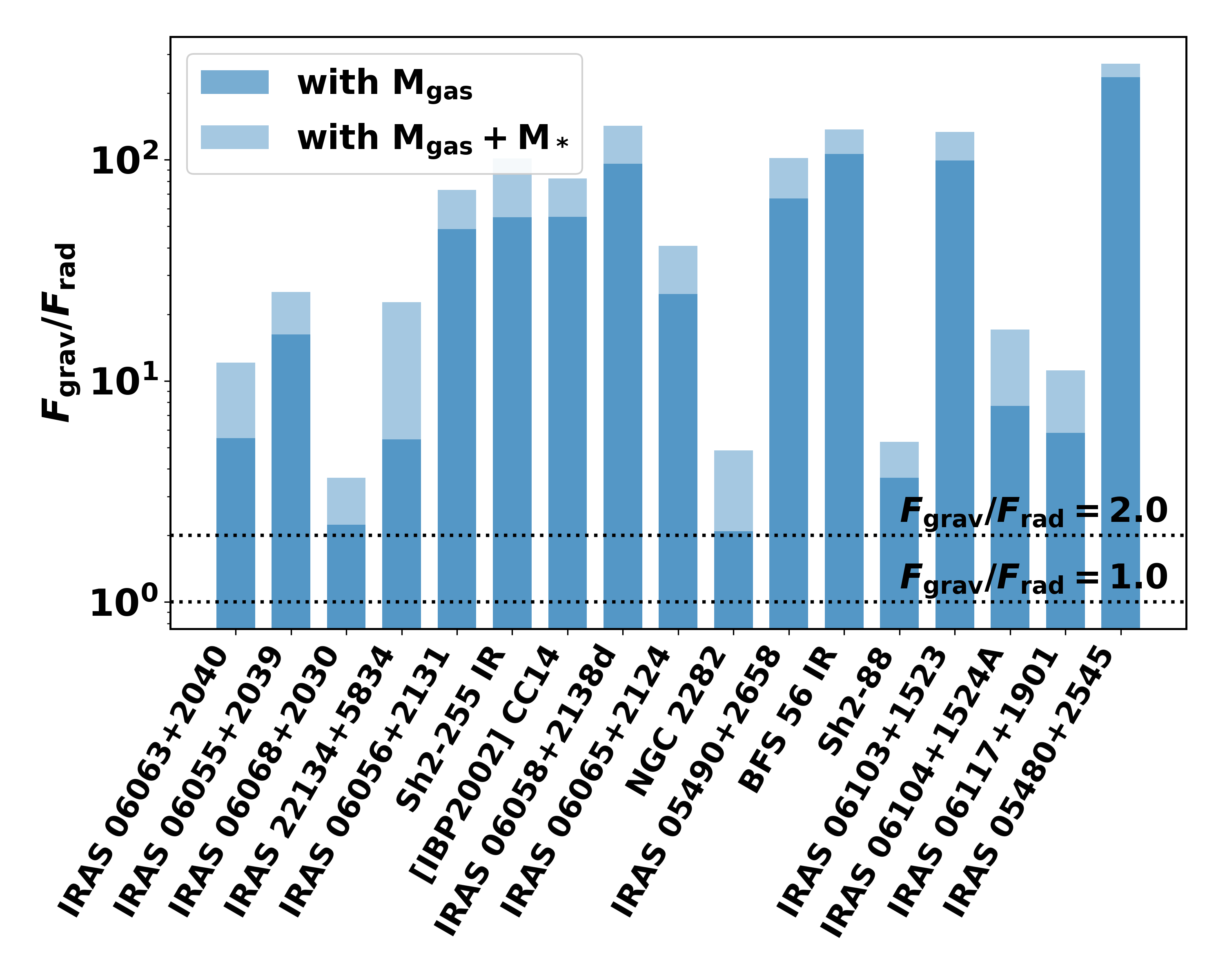}
    \caption{Ratio of gravitational to radiation pressure force for the clusters in our sample is shown. Dark blue bars represent values calculated using only the gas mass, while light blue bars include both gas and stellar mass in the calculation of $F_{\rm{grav}}$.}
    \label{Fig:Frad_Fgrav}
\end{figure}

As discussed above, the ionizing radiation from massive stars can affect the SFE of the clusters by halting the collapse or unbinding the gas present in the clump.  Therefore, it is important to investigate the relative strength of radiation pressure ($F_{\rm{rad}}$) and gravitational pressure ($F_{\rm{grav}}$) in the clumps. So, we calculated $F_{\rm{rad}}$ and $F_{\rm{grav}}$ using the estimated $N_{\rm{Lyc}}$, mass, and radius of the clumps, following the prescription given in \citet{mur09}
(for details, see Appendix \ref{Ionized}), and their ratios are given in Table \ref{tab:clusters:out}. Fig. \ref{Fig:Frad_Fgrav} shows the relative strength of both the forces in our sample, and as can be seen, the $\frac{F_{\rm{grav}}} {F_{\rm{rad}}}$ is far greater than 2 for majority of the clusters in our sample, suggesting that although \hii regions have developed in the 12 clumps, the clumps may still be collapsing and forming stars. As an example, we find that in the case of the Sh2-255 IR cluster in our sample, although NIR point sources and radio emissions have been observed by \cite{ojha11}, the region still hosts several compact millimetre cores with no infrared point sources \citep{zin12,zin18}, suggesting that stars are still forming in these clumps.

%shown in Fig. \ref{Fig:Frad_Fgrav}. The $\frac{F_{\rm{rad}}} {F_{\rm{grav}}}$ is given in Table \ref{tab:clusters:out}, which is greater than 2 for all the clusters. This shows that the clumps are still collapsing under gravity to form stars.}
%As discussed in Section \ref{sfe}, the depletion of gas material over time could lead to an overestimation of the SFE. Also, if massive stars have formed in a cluster, its radiation pressure tends to unbound the parent clump, which is more probable in relatively older clusters in our sample. Therefore, it is important to investigate the strength of radiation pressure ($F_{\rm{rad}}$) in comparison to the gravitational pressure ($F_{\rm{grav}}$) in the clumps. 
%The $F_{\rm{rad}}$ and $F_{\rm{grav}}$ are calculated using the $N_{\rm{Lyc}}$, mass, and radius of the clumps (for details, see Appendix \ref{Ionized}), which is shown in Fig. \ref{Fig:Frad_Fgrav}. The $\frac{F_{\rm{rad}}} {F_{\rm{grav}}}$ is given in Table \ref{tab:clusters:out}, which is greater than 2 for all the clusters. This shows that the clumps are still collapsing under gravity to form stars.}

\begin{figure}
    \centering
    \includegraphics[width= 8.5cm]{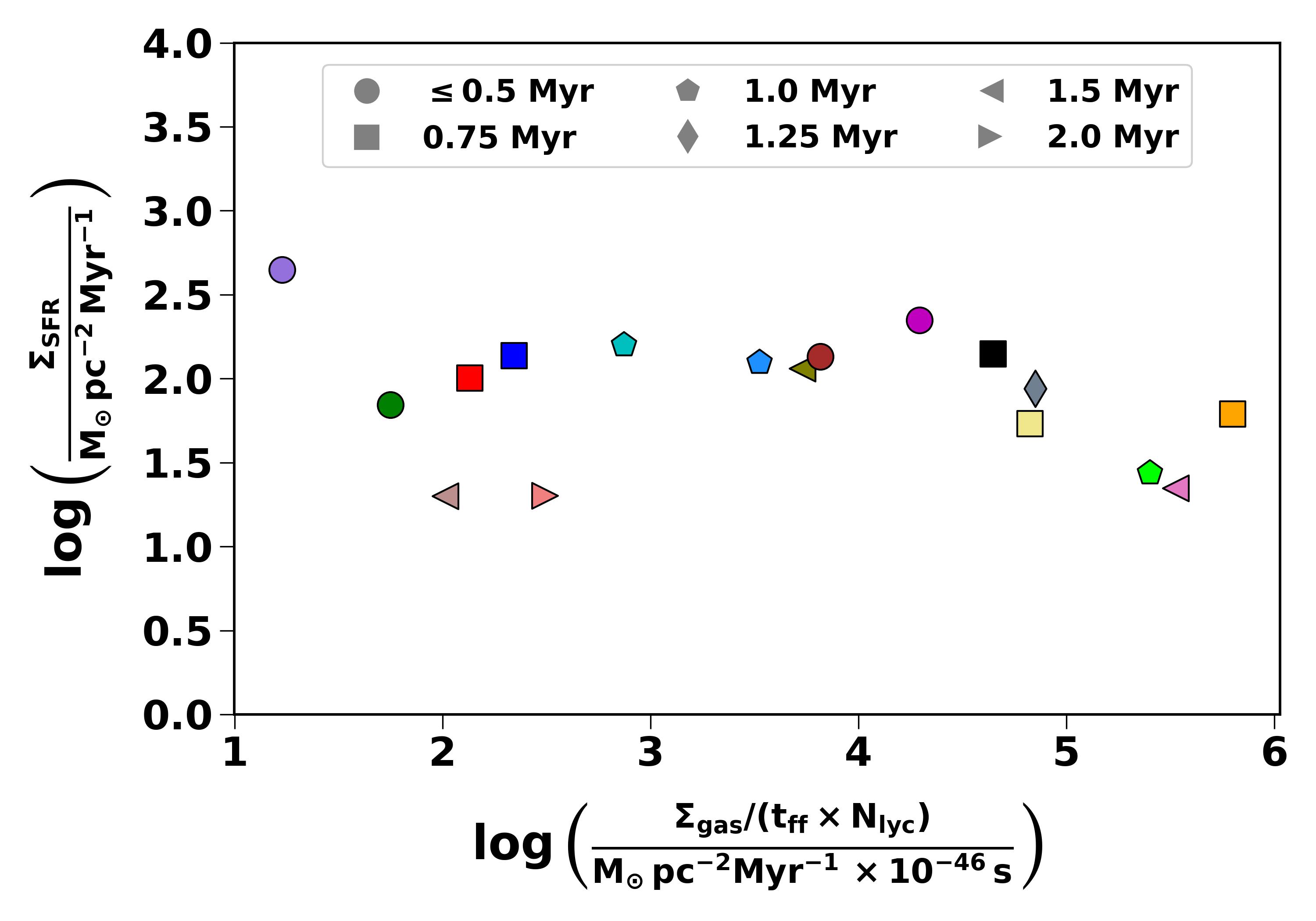}
    \caption{Comparison of $\log \Sigma_{\rm{SFR}}$ with $\log \Sigma_{\rm{gas}}/(t_{\rm{ff}} \times N_{\rm{Lyc}})$. The colored symbols are the same as in Fig. \ref{fig:SFR-gas_mass}.}
    \label{fig:scaling_Nlyc}
\end{figure}

%It is generally accepted that clouds with more massive stars are likely to terminate star formation earlier than clouds with less number of massive stars, unless there is a continuous supply of gas from an extended reservoir. 
%In our case, most of the clusters are located in isolated clumpy
%structures, so 
To understand the role of stellar feedback on the obtained scaling relations of the clumps, we normalized their gas mass surface densities by the associated number of Lyman continuum photons along with the free-fall time. Fig. \ref{fig:scaling_Nlyc} shows the variation of $\Sigma_{\rm{SFR}}$ with the $\frac{\Sigma_{\rm{gas}}} {t_{\rm{ff}}\, N_{\rm{Lyc}}}$ of the clusters. From the figure, it can be seen that $\Sigma_{\rm{SFR}}$ of the clusters remains flat with some
evidence of a declining trend at high Lyman continuum photons. Our results
hints to the fact that in young clusters like the ones studied in this work (i.e. regions of high gas surface densities), the stellar feedback does not seem to very significantly affect the star-formation properties. However, we acknowledge that given the small evolutionary range (i.e. 0.5$-$1.5 Myr for most of the sources) explored in this study and also the potential uncertainties associated with the age estimations, it is difficult to disentangle the effect of the evolutionary stages on the star formation properties of the studied sample. A better sample with a wider age range would be
valuable in this regard.
%The effects of feedback may become evident in the later evolutionary stages of the clumps.}
%Assuming that the narrow evolutionary sample has minimal effect on the
%star formation properties of the clump, our results seem to resonate with the findings of 
However, it is worth noting that, in a recent work, \cite{Zhou_2024c} found that feedback does not seem to alter the dynamics of the gas in protocluster-forming regions significantly.    

In summary, we believe that since our clumps are of compact size with high gas surface densities, currently actively forming stars, and the photoionization feedback effect is not yet significant, are
the likely cause of high SFEs and SFRs compared to other studies. 
Deep NIR and mid-IR observations (e.g. with JWST) of younger clumps, like the ones explored by \cite{Heyer_2016}, would shed more light on the effects of temporal evolution on the scaling relations at the clump scale.

\subsection{Implication on bound cluster formation and infant mortality}

%The clusters of mass around 10$^4$ \Ms~or more, i.e. young massive clusters (YMCs) are rare; for example, only 12 YMCs have been found in the Milky Way so far \citep[see Table 4 of][]{Krumholz_2019}, despite the fact that our Galaxy contains a large number of massive MCs ($>$ 10$^5$ \Ms). Thus formation of massive clusters like YMCs perhaps requires specific initial conditions, environment and mode of star formation in a given molecular cloud. The fact that makes the formation of massive clusters a difficult task is the very low SFE of 2$-$6\% that has been found in MCs \citep{Evans_2009, Lada_2010, Heiderman_2010, Evans_2014}. 
As discussed in Section \ref{chap_scaling_laws}, it is believed that the formation of bound clusters strongly depends upon the efficiency by which the gas in MCs or clumps gets converted into stars. Many simulations showed that the bound fractions of newborn stars are tightly correlated with the SFEs \cite[e.g.][]{shu17,li19, fukushima21,gurdic21}. Therefore, the SFE is a key parameter for understanding the formation and
early evolution of star clusters. Simulations suggest massive bound clusters ($\gtrsim$ 10$^4$ \Ms) formation through either a high mass gas assembly
with a high SFE ($\gtrsim$ 30\%) before the stellar feedback becomes significant \citep{Dib_2013, Longmore_2014, Ban_Kroupa_2015, Krumholz_2019}, or gradual gas assembly and hierarchical merger of small sub-clusters \citep{Longmore_2014, sema2019, Krumholz_2019, Polak_2023, Zhou_2025}.  
For example, \cite{David_2022} simulated a molecular cloud of mass $\sim$2 $\times$ 10$^4$ \Ms~and surface density $\sim$60 \Ms~pc$^{-2}$ and investigated its evolution over time by considering different physical factors like gas pressure, magnetic field, turbulence, and stellar feedback. The authors found that the cloud is able to form a cluster of mass $\sim$1.4 $\times$ 10$^3$ \Ms~with an efficiency of around 7\% in 4 Myr of time through the hierarchical assembly of gas, stars, and sub-clusters, adding relevance to the hierarchical scenario of cluster formation. At around 6 Myr, the feedback starts to affect the cloud significantly and disrupts the cloud in 8 Myr.  Although it is still possible to form bound clusters with low SFE, simulations in general suggest that massive bound clusters, in particular the young (less than a few Myr) and compact ones,  were most likely formed in regions with high SFEs \cite[e.g. see discussion in][]{fuji15,ban18}.
%because the later processes may take longer time.
%or a combination of both.
%In , we discussed the possible scenarios to form an intermediate-to-massive stellar cluster, i.e. either a high-gas mass reservoir with high SFE is needed, or a continuous supply of matter along with the merger of small subclusters or a combination of both \citep{Longmore_2014, Ban_Kroupa_2015, sema2019, Krum_2019}. 
%\textcolor{red}{xxxxxxxxxxxxxx.} %On the other hand, \cite{Polak_2023} in their simulations found that a molecular cloud of mass $\geq$ 10$^5$ \Ms~and surface density $\geq$ 100 \Ms~pc$^{-2}$ can form a bound cluster of mass $\sim$10$^4$ \Ms~with an efficiency of around 65\% in just one free-fall time. Also, the authors found that even a lower mass cloud is able to form a bound cluster but with a bound mass fraction of $\sim$60\%. 

For the studied cluster-forming clumps, the median SFE is found to be somewhat less than 0.3. As discussed above, most simulations suggest a high SFE ($\geq$ 30\%) is necessary to form massive bound clusters \citep{bau07,kru12,fuji15}. Thus, if these simulations are to be believed,  the low efficiency found in the studied clumps suggests that the clusters within them are unlikely to retain a significant fraction of bound stars as they age. However, we acknowledge that if the dense gas clumps are part of the larger cloud, they may remain gravitationally bound for a longer time with the continuous supply of matter and may continue to form stars.
%\textcolor{red}{into open clusters, given that most clusters are located in isolated clumps or are only weakly connected to the extended gas reservoir.} 
Although stellar feedback is not yet significant in these clusters, the fact that the median disk lifetime of young stars in young clusters is found to be around 2.5 Myr \citep[see discussion in][and references therein]{Rawat_2024c}, implies that the formation of young stars declines significantly beyond 2.5 Myr. Thus, we do not expect a significant rise in the SFE in most of these clusters, although simulations with characteristics of the studied sample are needed to ascertain this hypothesis. Nevertheless, our results points to the fact that the low SFE of the cluster forming clumps in the Galactic plane could be a possible reason for the "infant mortality" inferred from the statistics of embedded to open clusters by \citet{lada_lada_2003}, where a high fraction of young clusters/protoclusters dissolve in the Galactic field, and only a few per cent remain bound. Our results suggest that the formation of massive bound clusters likely requires a special environment, such as the high-pressure conditions expected such as in galactic center or in galaxy mergers or colliding substructures that compress a larger fraction of gas over short timescales \cite[e.g. see discussion in][]{kru25}, rather than the environment studied in this work.

%In chapter \ref{chap_dust} and \ref{chap_gas}, we present and discuss the possibility of hierarchical assembly of matter along with the global collapse of the \cloud~cloud to form a stellar cluster. We discussed that the cluster, which is forming at the hub, might grow to become at least an intermediate-mass cluster if it continues to form stars at the same rate along with a continuous supply of matter from the filaments. In this chapter, we explored the second possibility, i.e. a high SFE in a dense environment (clumps). A mean SFE of $\sim$0.27 is found for the cluster forming clumps in our sample. Most of the clusters are at their early stages of star formation, i.e. the median age in our sample is around 1.2 Myr, which means that they can further grow in mass in a few Myrs. However, to form even an intermediate-mass cluster, they would require a high-mass gas reservoir, which is at present available for only two-three clusters. Nevertheless, a combination of both the aforementioned scenarios of cluster formation is also possible, i.e. a hierarchical assembly of matter from the surrounding to the protocluster forming in the clump, and the clump is the highly dense region forming the stars with an SFE of around 20$-$30\%. With a gradual supply of gas from the natal cloud, an intermediate-to-massive cluster may form. We also found evidence of this scenario in our study of \cloud~in which a young cluster, FSR 655, is forming at the central clump/hub of the cloud, while the clump is also accreting the gas from the surrounding through filaments.          

\section{Summary and Concluding Remarks}
\label{sec:summary}
We present the statistical analysis of a sample of 17 cluster-forming clumps to investigate the relation between star formation rate and gas surface density at the clump scale. We constrained our sample up to a distance of 2.2 kpc for better characterization of cluster properties. The UKIDSS NIR photometric data and $\it{Herschel}$ dust continuum-based column density maps were used to derive various properties of the clusters, such as extinction, age, stellar and gas mass, SFR, and SFE. The SFRs in our sample range from $\sim$29 to 500 \Ms~Myr$^{-1}$, with a mean around 184 \Ms~Myr$^{-1}$. The instantaneous SFEs range from $\sim$0.07 to 0.51, with a mean around 0.23. The mean and median $\Sigma_{\rm{gas}}$ of the clumps is $\sim$314 \Ms~pc$^{-2}$ and $\sim$258 \Ms~pc$^{-2}$, respectively.

We found that $\Sigma_{\rm{SFR}}$ varies with $\Sigma_{\rm{gas}}$ as $\Sigma_{\rm{SFR}} \propto \Sigma_{\rm{gas}}^{(1.46 \pm 0.28)}$ in the studied sample of cluster-forming clumps, and both quantities are well correlated. The $\Sigma_{\rm{SFR}}-\Sigma_{\rm{gas}}$ relation in this work lies well above most of the previously obtained relations at extragalactic, cloud, and clump scales.
%which might be due to the fact that we have studied the scaling relations in compact clumps that are associated with embedded star clusters. 
%and chosen only the active star-forming clumps. 
The volumetric star formation relation is found to be of the form $\Sigma_{\rm{SFR}} \propto ({\Sigma_{\rm{gas}}/t_{\rm{ff}}})^{(0.80 \pm 0.15)}$, and it also lies above the previously obtained relations at the cloud scale. The mean $\epsilon_{\rm{ff}}$ of clumps found in this work is around 15\%, which is significantly higher than the constant efficiency per free-fall time reported for nearby MCs \citep{Krum_mckee_2005, Krumholz_2012}.  We suggest that the high values observed in the $\Sigma_{\rm{SFR}}$ vs. $\Sigma_{\rm{gas}}$ and $\Sigma_{\rm{SFR}}$ vs. $\Sigma_{\rm{gas}}/t_{\rm{ff}}$ planes for the studied clumps are likely due to the fact that we have studied the scaling relations in compact clumps ($<$1.6 pc) with high gas surface densities that are associated with active embedded star clusters. This also implies that star formation scaling relations may not be universal and depend on local conditions, scales, and evolutionary status. Simulations also suggest a higher star-formation rate per free-fall time in compact and dense clouds
than fiducial clouds \citep[e.g.][]{he19}. Most of the clumps in our sample have $\Sigma_{\rm{gas}}$ $\gtrsim$ 110~\Ms~pc$^{-2}$, and the SFR$-$gas mass relations show a good correlation, which favours the conclusions of \cite{Lada_2010}, \cite{Heiderman_2010}, and \cite{Evans_2014}, that the SFR$-$gas mass relations become better correlated at higher threshold density. 

We find that the median SFE of the cluster-forming clumps is somewhat lower than the expected high SFE required for star clusters
to remain bound as they age, as suggested by simulations. Thus, the studied clusters might not survive violent gas expulsions to remain bound for a longer time. %{\bf We suggest that observed low SFE could be one of the potential reasons of the infant mortality of star clusters.} 

Overall, our results do not favour a universal relation between star formation rate and gas mass that can explain the star formation process 
across all scales, from galaxies and giant molecular clouds to smaller structures like clouds and clumps.

\section*{ACKNOWLEDGEMENT}

%We thank the anonymous referee for the comments and suggestions that helped to improve the paper. 
We thank the anonymous referee for the useful comments and suggestions to improve the quality of the paper. The research work at the Physical Research Laboratory is funded by the Department of Space, Government of India. D.K.O. acknowledges the support of the Department of Atomic Energy, Government of India, under Project Identification No. RTI 4002. This work makes use of data obtained from the UKIRT Infrared Deep Sky Survey, obtained using the wide field camera on the United Kingdom Infrared Telescope on Mauna Kea. We acknowledge the Herschel Hi-GAL survey team and
ViaLactea project funded by EU carried out at Cardiff University. We thank Matthew W. Hosek Jr. for all the discussions and support related to the SPISEA code. We are also thankful to Mark Heyer for sharing the data from his work on massive clumps to compare with this work.

\section*{Data Availability}

For this study, we have used the NIR photometric data from UKIDSS, which is available on the WFCAM science archive, \href{http://wsa.roe.ac.uk/}{http://wsa.roe.ac.uk/}. We have also used the FIR data from $\it{Herschel}$, which is available on the archive. The $\it{Herschel}$ column density maps based on PPMAP are available on the archive, \href{http://www.astro.cardiff.ac.uk/research/ViaLactea/PPMAP_Results/}{http://www.astro.cardiff.ac.uk/research/ViaLactea/PPMAP\_Results/}. The radio data from the NVSS survey used in this work is also available on the archive, \href{https://www.vla.nrao.edu/astro/nvas}{https://www.vla.nrao.edu/astro/nvas}.



\bibliographystyle{mnras}
\bibliography{myref.bib} % if your bibtex file is called example.bib
%\begin{thebibliography}{}
%\end{thebibliography}

%%%%%%%%%%%%%%%%%%%%%%%%%%%%%%%%%%%%%%%%%%%%%%%%%%

%%%%%%%%%%%%%%%%% APPENDICES %%%%%%%%%%%%%%%%%%%%%

\appendix

\section{Ionized gas properties of the clumps}
\label{Ionized}

\begin{figure}
    \centering
    \includegraphics[width= 8.5cm]{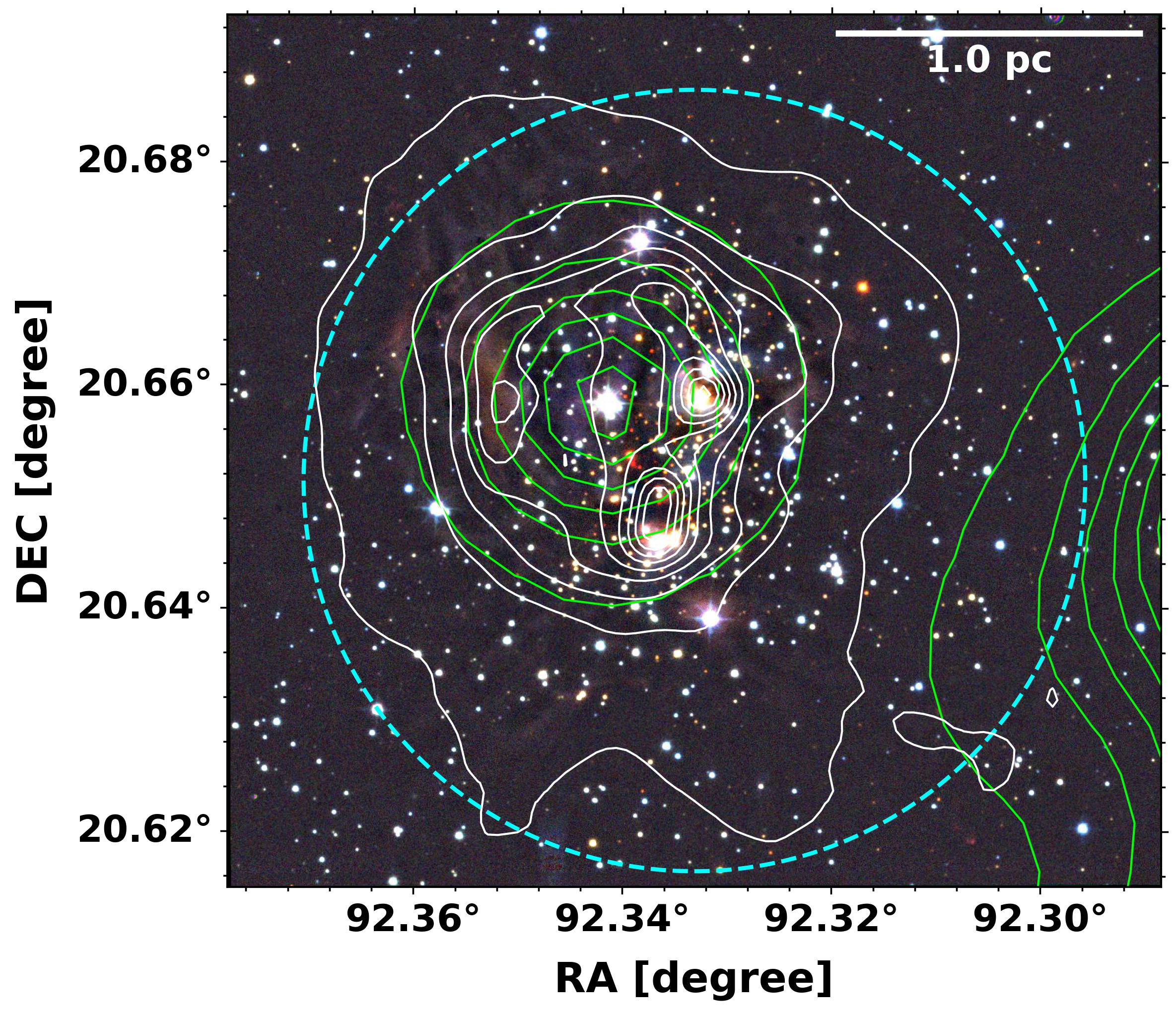}
    \caption{UKIDSS color-composite image of IRAS 06063$+$2040 over which the contours of ionized emission at 1.46 GHz and 70 $\mu$m emission are shown by green and white colors, respectively. The contour levels are shown above 5$\sigma$ from the background.}
    \label{fig:iras_2040_radio}
\end{figure}

Fig. \ref{fig:iras_2040_radio} shows the 1.46 GHz radio emission and 70 $\mu$m emission over the UKIDSS image of IRAS 06063$+$2040. To calculate $N_{\rm{Lyc}}$ from $S_{\rm{\nu}}$, we used the following relation \citep{rub68}

\begin{equation}
    N_{\rm{Lyc}} = 4.76 \times 10^{48} \left(\frac{S_{\rm{\nu}}}{\rm{Jy}}\right)  \left(\frac{T_{\rm{e}}}{\rm{K}}\right)^{-0.45}  \left(\frac{\nu}{\rm{GHz}}\right)^{0.1} \left(\frac{d}{\rm{kpc}}\right)^2,
\end{equation}
where $T_{\rm{e}}$ is the electron temperature, $\nu$ is the frequency of the radio emission, and $d$ is the distance to the clumps. We adopted an average value of $T_{\rm{e}}$ to be around 8200 $\pm$ 240 K from \cite{Quireza_2006}. Using $N_{\rm{Lyc}}$ and the masses and sizes of the clumps, the radiation and gravitational pressure forces for the clumps can be derived as \citep{mur09} 

\begin{equation}
    F_{\rm{rad}} = 5 \times 10^{34} \left(\frac{L}{4 \times 10^{11} \lsun} \right) \rm{dyn},
\end{equation}

\begin{equation}
    F_{\rm{grav}} = 3 \times 10^{34} \left(\frac{M}{10^9 \rm{\Ms}}\right)^2 \left(\frac{1 \rm{kpc}}{R} \right)^2 \rm{dyn},
\end{equation}
where $L$ is the luminosity of the ionizing source. We have taken the luminosity values corresponding to the Lyman continuum photons given in \cite{Thompson_1984}.

\section{Age determination from SED modelling}
\label{SED}
\begin{figure}
    \centering
    \includegraphics[width=8.55cm]{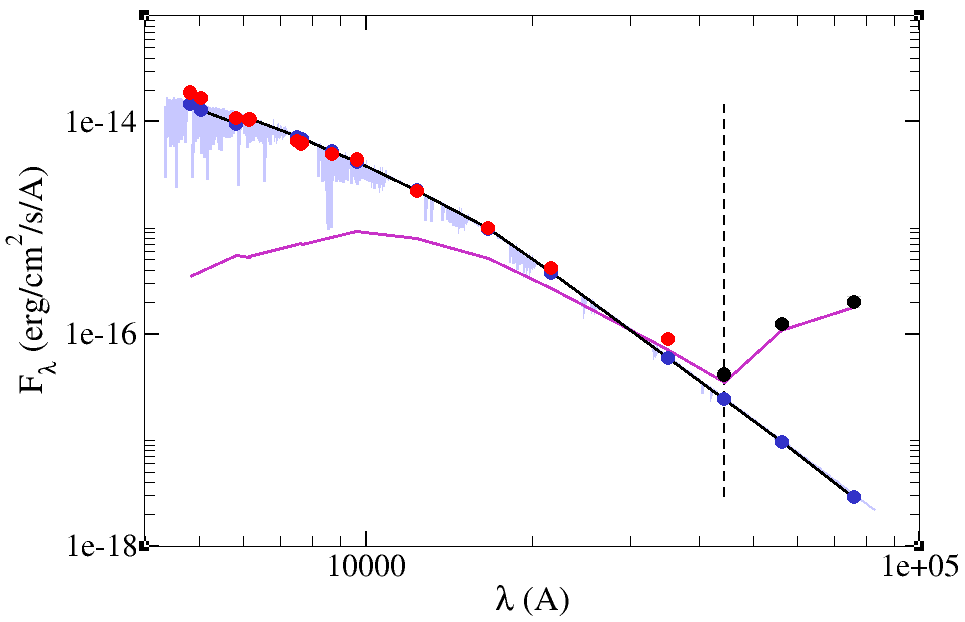}
    \caption{Example of a SED fitting as generated by VOSA for IRAS 06063$+$2040. The blue spectrum represents the theoretical model that fits best, while red dots represent the observed photometry. The black dots are affected by IR excess and are not considered in fitting. 
}
    \label{fig_vosa}
\end{figure}

In order to validate the age estimation of the clusters from KLF modelling, we accessed the age of IRAS 06063$+$2040 using SED modelling. We used VOSA \citep[VO Sed Analyzer;][]{Bayo_2008} tool, which has been effectively used to derive the stellar properties of young stars \cite[e.g.][]{muzic19, almen23, gupta24}. VOSA compares the observed fluxes with the synthetic photometry, looking for the best-fit effective temperature ($T_{\rm{eff}}$), extinction ($A_{\rm{V}}$), and surface gravity (log(g)) combination. For fitting, in addition to UKDISS data, we used optical to NIR data from GAIA DR3, Pan-STARRS,  $\it{Spitzer}$, and WISE, as available in the Vizier. We used the distance 2.2 $\pm$ 0.2 kpc, the BT-Settle atmospheric models \citep{Allard_2014}, $T_{\rm{eff}}$ over the range of 1000$-$7000 K, and \av as 6 $\pm$ 3 mag. For objects showing excess in $\it{Spitzer}$ photometry, SED fitting is performed over the shorter wavelengths; otherwise, the full available range is included in the fit. We used the log(g) between 3.5 and 5.0, which is suitable for young low-mass stars and the field stars; however, we note that the VOSA SED fitting procedure is largely insensitive to log(g) \citep[see][]{Bayo_2017}. In fact, we find that, in general, the fits are not significantly affected by variations in log(g). With the aforementioned approach, we derived effective temperatures and luminosities for the sources. The radii of our objects are calculated using the Stefan-Boltzmann equation, making use of the luminosities and effective temperatures previously computed by VOSA.
With the derived stellar parameters, VOSA also provides the age of the sources based on their position on the HR diagram using theoretical isochrones. To determine the age of the cluster, we keep only those sources with good SED fitting, i.e. sources with Vgf$_b$ value less than 15. Vgf$_b$ is a modified reduced $\chi^2$, calculated by forcing that the observational errors are, at least, 10\% of the observed flux. In general, Vgf$_b$ $<$ 15 is usually considered as a proxy for well-fitted SEDs \cite[e.g.][]{reba21,nayak24}. We further remove the foreground sources from the sample by selecting sources whose $J-H$ and $H-K$ colors are less than 0.8 and 0.25 mags, respectively. To identify foreground sources, we compared the NIR color-color diagram of the cluster population with that of the control field population, as described in \cite{Rawat_2024c}, and selected sources based on this comparative analysis. With the aforementioned criteria, we are left with only 15 sources, whose median age was found to be 0.8 $\pm$ 0.4 Myr.  Figure \ref{fig_vosa} shows an exemplary SED  fitted with VOSA. 

%We find this age, within the error, is in reasonable agreement with the age derived based on the KLF method. 

%{\bf Kusagra need to verify the above facts}

%\begin{figure*}
   % \centering
    %\includegraphics[width= 16cm]%{SFR_gas_mass_GHC_comparison.jpeg}
    %\caption{Comparison of $\Sigma_{\rm{SFR}}-\Sigma_{\rm{gas}}$ relation obtained in this work with the GHC model. The black solid curve shows the evolution of $\Sigma_{\rm{SFR}}$ and $\Sigma_{\rm{gas}}$ of a 2000 \Ms~modeled cloud \citep[for details, see][]{Zamora_2012}.}
    %\label{fig:sfe_GHC}
%\end{figure*}

%%%%%%%%%%%%%%%%%%%%%%%%%%%%%%%%%%%%%%%%%%%%%%%%%%

% Don't change these lines
\bsp	% typesetting comment
\label{lastpage}

\end{document}